\documentclass{aa}  

\usepackage{graphicx}
\usepackage{txfonts}
\begin{document}

   \title{Spot Modeling through Multiband Photometry}

   \subtitle{Analysis of V1298 Tau}

   \authorrunning{A. Biagini et al. }
   \author{Alfredo Biagini
          \inst{1} \inst{2},
          Antonino Petralia \inst{1},
          Claudia Di Maio \inst{1},
          Lorenzo Betti \inst{3} \inst{4},
          Emanuele Pace \inst{3} \inst{4}
          \and
          Giuseppina Micela\inst{1}
          }
   
   \institute{INAF - Osservatorio Astronomico di Palermo, Piazza del Parlamento 1, 90134 Palermo, Italy
         \and
             Università degli Studi di Palermo, Dipartimento di Fisica e Chimica, Via Archirafi 36, Palermo, Italy\\
              \email{alfredo.biagini@inaf.it}
         \and
             Dipartimento di Fisica ed Astronomia, Università degli Studi di Firenze, Via Sansone, 150019 Sesto Fiorentino (FI), Italy   
         \and
             Osservatorio Polifunzionale del Chianti, Barberino Val d’Elsa, Florence, Italy}

   \date{}

 
  \abstract
   {Stellar activity consists of different phenomena, mainly spots and faculae, and it is one of the main sources of noise in exoplanetary observations because it affects both spectroscopic and photometric observations. If we want to study young active planetary systems we need to model the activity of the host stars in order to remove astrophysical noise from our observational data. }
   {We modelled the contribution of stellar spots in photometric observations. Through the use of multiband photometry, we aim to extract the geometric properties of the spots and constrain their temperature.}
   {We analyzed multiband photometric observations acquired with the 80 cm Marcon telescope of the Osservatorio Polifunzionale del Chianti of V1298 Tau, assuming that the photometric modulation observed in different bands should be due to cold spots.}
   {We constrained the effective temperature of the active regions present on the surface of V1298 Tau, which is composed by the contemporary presence of spots and faculae. We tested our hypothesis on solar data, verifying that we measure the size of the dominant active region and its averaged effective temperature.}
   {}


   \maketitle
%

\section{Introduction}

Stellar variability involves a variety of inhomogeneities on the stellar surface, produced by the magnetic activity of the interior of the star \citep{2005LRSP....2....8B} that leads to a large variety of phenomena like stellar spots \citep{2003A&ARv..11..153S}, faculae and flares \citep{2012A&A...539A.140B} and their intrinsic evolution with a large variety of timescales:  minutes for flares, days for the evolution of the active regions (faculae and spots) on the stellar surface, years for active regions cycles, billions of years for the evolution of the magnetic activity of the star \citep{2005LRSP....2....8B}.\\
The analysis of the stellar activity is crucial for understanding the stellar interior processes. Moreover, it is also fundamental for exoplanetary observations, since it is the main source of noise both for photometry and spectroscopy \citep[e.g.][]{2008MNRAS.385..109P,2009A&A...505.1277C,2009A&A...505..891S,2011MNRAS.416.1443S,2011A&A...527A..73S,2010ApJ...721.1861A,2011A&A...526A..12D,2011ApJ...736...12B,2012A&A...539A.140B,2015ExA....40..723M,2015ExA....40..711S}.\\
Stellar spots are regions with a lower temperature with respect to the quiet stellar surface because the local magnetic field configuration halts the convection process leading to a local cooling of the stellar plasma together with a change in local gravity and magnetic field \citep{2005LRSP....2....8B,2012A&A...539A.140B}. 
The presence of stellar spots, for instance, can influence both spectroscopic and photometric observations: it can distort spectral lines, thus hampering accurate radial velocity estimation \citep{2008MNRAS.385..109P,2020A&A...638A...5C,2023A&A...672A.126D} or affect the photometric curves analysis by hiding or altering the signal of a transit in stellar light curves \citep{2018AJ....155..156T}.
Furthermore, stellar activity could impact the extraction of the exoplanetary transmission spectra\citep{2020AJ....160..260C}, hampering the detection of atmospheric molecules and atmospheric features like clouds \citep{2008MNRAS.385..109P,2012A&A...539A.140B,2015ExA....40..723M}.\\
The modelling of the stellar inhomogeneities on the stellar surface could play a crucial role in the search and characterization of exoplanetary atmosphere for mission such as JWST \citep[][]{2016ApJ...817...17G} and for the incoming ARIEL mission \citep[Atmospheric Remote-Sensing Infrared Exoplanet Large-survey,][]{2018ExA....46..135T}, the M4-ESA mission dedicated to study the exoplanetary atmospheres.\\
Stellar activity is a very common phenomenon among solar-type stars and it is due to the effects of the magnetic dynamo inside the star \citep{2005LRSP....2....8B}. Because stellar magnetic dynamos increase along with the stellar rotation \citep{1981ApJ...248..279P,1981ApJ...245..671W,2003A&A...397..147P,2005LRSP....2....8B} and due to the typical age-rotation relation \citep{1995ApJ...438..269B}, we find that stellar activity is particularly strong for young stars \citep{1995ApJ...438..269B,2005LRSP....2....8B}. Despite this, in recent years, young stars have become targets of significant interest because they can provide valuable insights into the early stages of formation and evolution of exoplanetary systems \citep{2016AJ....151..112D,2020A&A...638A...5C,2020AJ....160...33R,2020Natur.582..497P,2023A&A...672A.126D,2024A&A...682A.129M}.\\
In recent years, ground- and space-based facilities have enabled us to discover thousands of exoplanets orbiting various types of stars, revealing a variety of these bodies with characteristics different from those of our own Solar System.\\
In order to determine the properties of young planetary systems  and analyze their planetary atmospheres, it is fundamental to understand the stellar activity of the host stars to remove its effects from the observational data or to study its impact on planets.\\
In this paper, we focus our work on spots analysis. In particular, we study only the change in temperature in the spots, with the hypothesis that local gravity and magnetic field changes of the spot have much more little effects on stellar observations.\\
Due to their lower temperature, the spot emission curve differs from that of the quiet photosphere, thereby altering the photometric signal, with an effect that decreases along the wavelength observed, or distorting the spectral lines of the star and leading to incorrect radial velocity measurements \citep{2005LRSP....2....8B}.
For instance, during a photometric observation, the presence of a spot occulted by a transiting planet can produce bumps in the observed light-curve. This occurs because the planet obscures a region with a lower flux compared to the unspotted surface, thereby increasing the observed total intensity coming from the star  \citep{2009A&A...505.1277C,2011MNRAS.416.1443S}.\\
On the other hand, also an unocculted spot can affect the measurement of the transit depth  \citep{2009A&A...505.1277C} since it can affect both photometric and spectroscopic in-transit observations without clear signs of its presence, leading to incorrect estimations of the exoplanets parameters. In particular, the chromaticity of these effects leads to inaccurate measures of the planetary radius at different wavelengths.  The presence of a spot can also mimic a transit and, as a consequence, more than one observations are needed to confirm the presence of a transiting planet. \\
In order to remove the effects of the spots on our data, we need to know their properties, positions and temperature. For this reason, in the following, we will develop a method to model the spot configuration in active stars based on a multiband photometry analysis.
In this work, we used multiband photometric ground-based observations to model the spots of a very interesting young stellar object (YSO), V1298 Tau, that we will describe in Section 2.
In Section 3 we will describe the methodology of our observations and in Section 4 our retrieval method, while in Section 5 we will discuss the validation of our model using solar data.
Our final results will be discussed in Section 6.

\section{Stellar target: V1298 Tau}
V1298 Tau is a K0-K1.5 young star ($\mathrm \sim$20 Myr) with a mass about $\mathrm{1.1 M_{\odot}}$ located at $\mathrm \sim$ 108 pc from Earth. It is an ultra-fast rotator with a rotational period of less than 3 days \citep{2019AJ....158...79D}. In Table \ref{V1298_properties} we show the properties of V1298 Tau from the literature.\\
V1298 Tau is one of the youngest stars known to host more than one transiting planet \citep{2022NatAs...6..232S}, not a very common occurrence \citep{2011ApJ...732L..24L}.
The star and its planets have already been studied both through photometry \citep[for example by TESS,][]{2022ApJ...925L...2F} and spectroscopic observations \citep[through radial velocities, for example by HARPS-N, ][]{2022NatAs...6..232S}.\\
We know also from previous studies, such as the spectral cross-correlation (CCF) function analysis from \cite{2023arXiv231214269D}, that the star shows large spots at high latitudes (> 60°) and smaller spots at low latitudes (< 40°), but the relation between the spectral distortion due to stellar activity and the photometric flux variations of the star is not clear and the spectral analysis of the spots has been based on the hypothesis of not emitting spots (i.e. black spots, surface regions with temperature of 0 K) that is not physically correct.
In this paper, we are interested in modelling the activity of the star due to spots on the stellar surface using multiband photometric data to better constrain their temperature and geometric parameters.

\begin{table}[!h]
\caption{\label{V1298_properties} V1298 Tau properties \citep{2019AJ....158...79D,2019ApJ...885L..12D}.}
\centering
\begin{tabular}{ l | c }
\hline
\hline
\rule{0pt}{4ex} \textbf{Parameter} & \textbf{Value} \\
\hline
\rule{0pt}{4ex}$\mathrm{M_{*}  (M_{\odot})}$&  $1.101\pm 0.005 $\\
\rule{0pt}{4ex}$\mathrm{R_{*}  (R_{\odot})}$& $1.34\pm 0.06$\\
\rule{0pt}{4ex}Age (Myr)  & $23\pm4$\\
\rule{0pt}{4ex}$\mathrm{P_{rot}}$ (d) &$2.87\pm 0.02$\\
\rule{0pt}{4ex}T (K)& $ 4970\pm 120$ \\
\rule{0pt}{4ex}v sin i $\mathrm{(km\: s^{-1})}$ &$ 23 \pm 2$\\
\rule{0pt}{4ex}Distance (pc)&$\mathrm{108.5 \pm 0.7}$\\
\rule{0pt}{4ex}Spectral Type &  $\mathrm{K0-K1.5}$\\
\rule{0pt}{4ex}V mag & $10.12$\\
\hline
\end{tabular}
\end{table}

\begin{table}[!h]
\caption{\label{SETS} Scheme of our observing campaign of V1298 Tau. OPC is "Osservatorio Polifunzionale del Chianti".}
\centering
\begin{tabular}{ c | c  }
\hline
\hline
\rule{0pt}{3ex}\textbf{Time}  & \textbf{Bands}\\
\hline
\rule{0pt}{4ex} 21-25/02/2021&B-V-R\\
\hline
\rule{0pt}{4ex} 11-15/12/2021&B-V-R\\
\hline
\rule{0pt}{4ex} 21-23/02/2022& B-V-R-I\\
\hline
\rule{0pt}{4ex} 28/02-01-02/03/2022& B-V-R-I\\
\hline
\end{tabular}
\end{table}

\section{Observations and data analysis}
Using the photometric light curve to model the spots leads to the so-called inverse light curve inversion problem \citep{2021AJ....162..123L}: we can't see directly the spots, so we have many geometrical degeneracies in our models. For example, it is difficult to distinguish spots on each specific hemisphere of the star and we have a degeneracy between temperature, radius and latitude of the spots. For isolated observations, we have also a degeneracy between the latitude and longitude of the spots that can be broken by observing the star multiple times during its rotation.
We aim to break these degeneracies by monitoring the star for a few days through different photometric bands at the same time. We chose to use multiband photometric observations because we want to take advantage of the chromaticity of the spot contribution to the stellar observation: we expect in fact to have a different ratio between spots flux and stellar flux at different wavelengths because they correspond in first approximation to two blackbodies with different temperatures. The difference between the spots and stellar flux should decrease at longer wavelengths, as far as in IR band: for example, a star with $\mathrm{T_{\odot}}$  and a spot with a filling factor of $1\%$ and a $\mathrm{\Delta T}$ of $\mathrm{1250 \ K}$ will present variations of  $\mathrm{\frac{\Delta FU}{FU} =  9\times10^{-3} }$ and $\mathrm{\frac{\Delta FK}{FK} = 3\times10^{-3}}$, respectively \citep{2012A&A...539A.140B}. \\ 
Through simultaneous multi-band photometric observations, we aim to estimate the temperature difference between the stellar surface (whose temperature is known from literature) and the spots.\\
\subsection{Observations campaigns}
We have organized multiband photometric observing campaigns from Osservatorio Polifunzionale del Chianti (OPC) with a Ritchey-Chretien telescope with the following characteristics:
\begin{itemize}
    \item \rule{0pt}{3ex} 80 cm diameter
    \item \rule{0pt}{3ex} f/8
    \item \rule{0pt}{3ex} field of view (FoV) $\mathrm \sim$ 20’x20’
    \item \rule{0pt}{3ex} Johnson filters B-V-R-I
\end{itemize}
Each observing campaign spanned between 3 and 5 days, i.e. between 1 and slightly less than 2 rotational periods (about 2.8 days) of the star (see Table \ref{SETS}).
The star's location near the ecliptic during the period from October to March poses challenges in coordinating observations for this target. Thus, in addition to the usual constraints arising from potential bad weather conditions, we also had to account for the lunar interference by avoiding days when the Moon passed near the target within a 60° range.
Finally, to model the stellar photosphere, we avoided all the days with planetary transits for this target.
In conclusion, we observed the star in three different periods: in February 2021, December 2021 and February 2022. 
We evaluated the best observing runs selecting only observed sets with $\mathrm{SNR>20}$. Finally, to avoid problems due to possible spot evolution we limited our analysis to maximum 3 days (see Section 4), obtaining these final databases:
\begin{itemize}
    \item \rule{0pt}{3ex}21-22-23, 22-23-24 and 23-24-25 (February 2021) with B, V and R bands
    \item\rule{0pt}{3ex}12-13-14 and 13-14-15 (December 2021) with B, V and R bands
    \item\rule{0pt}{3ex}11-12-13 (December 2021) with V and R bands
    \item\rule{0pt}{3ex}21-22-23 (February 2022) with V, R and I bands.
\end{itemize}

\begin{figure*}[!t]
		\centering
            \includegraphics[scale=0.4]{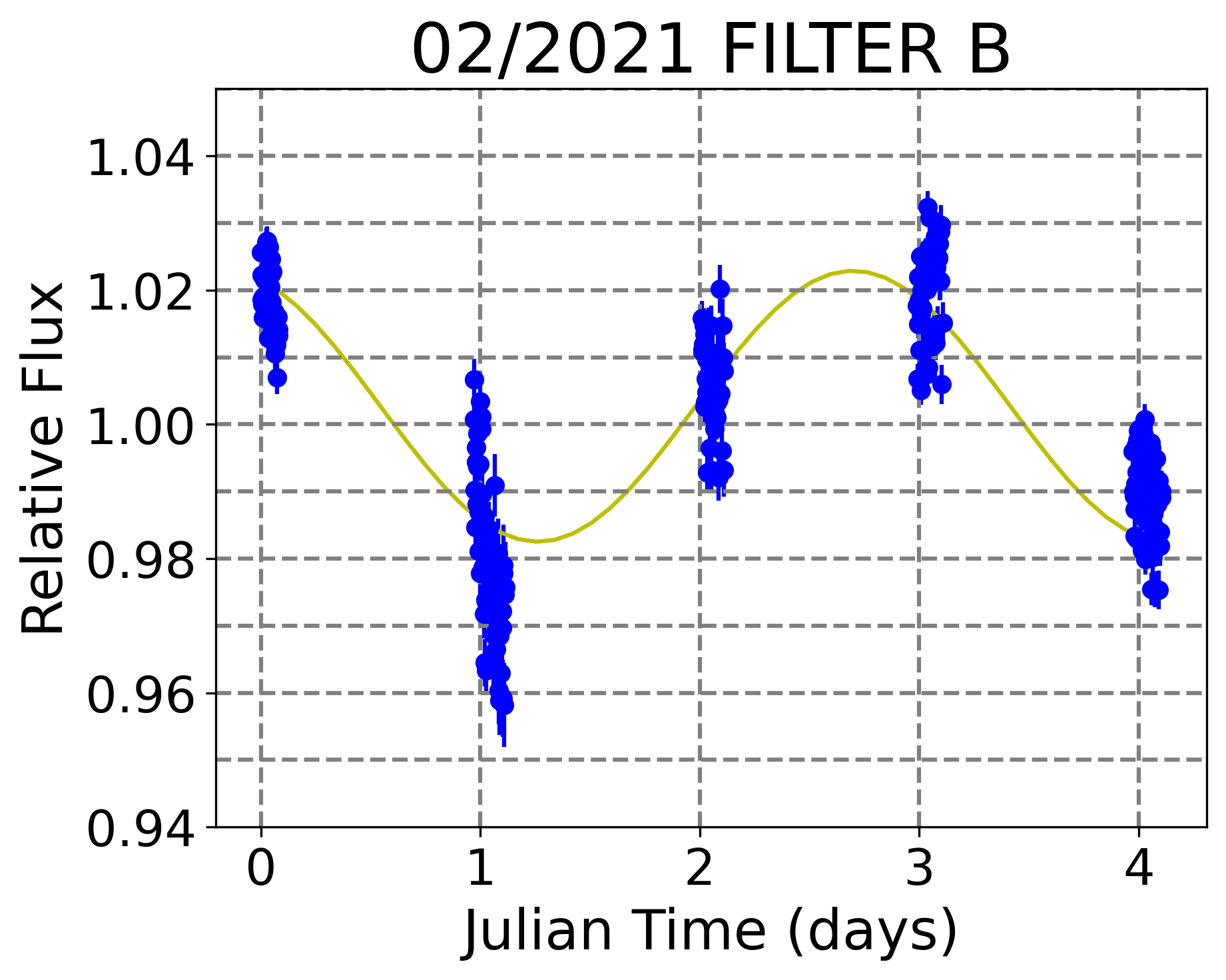}
            \includegraphics[scale=0.4]{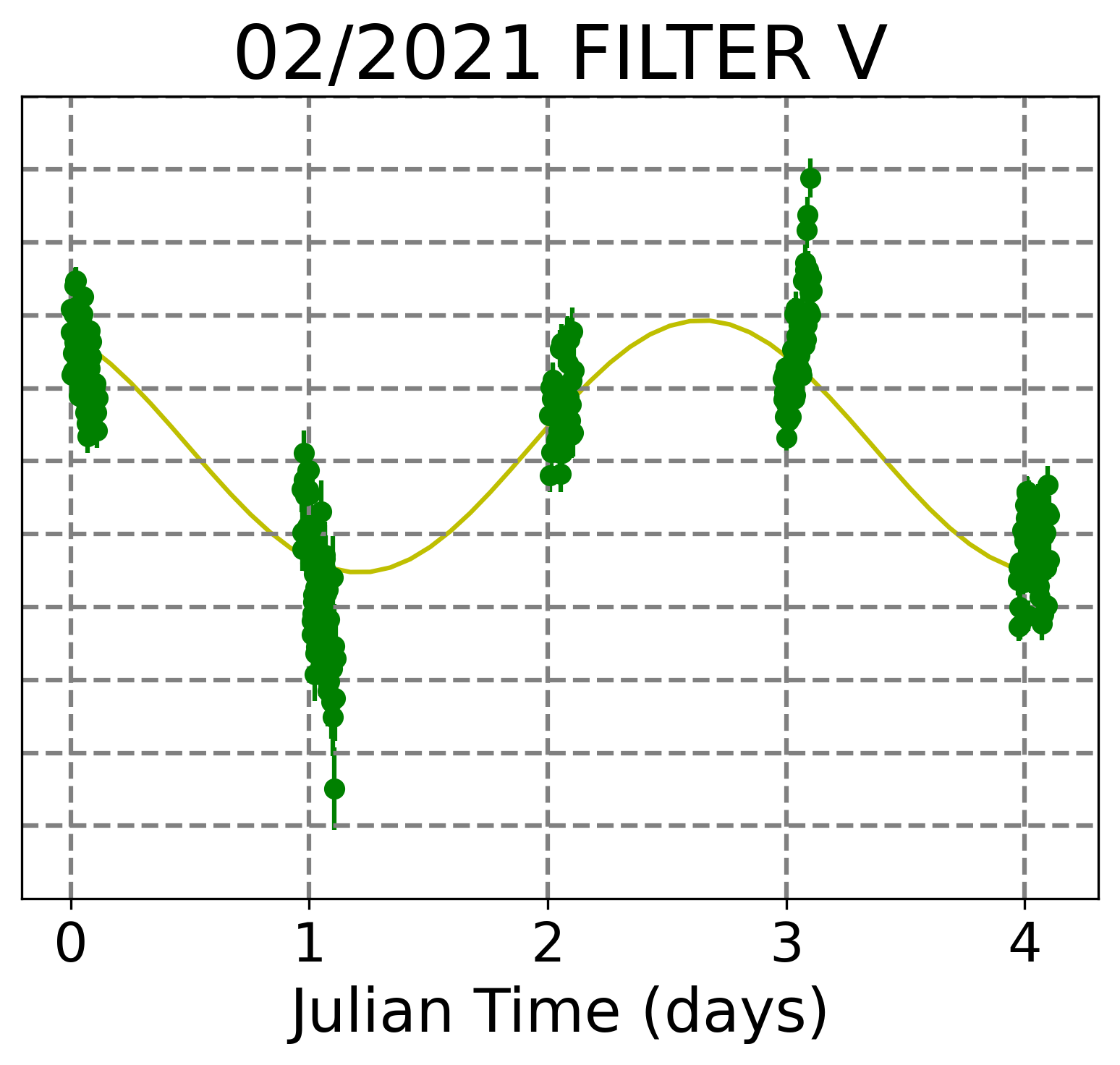}
            \includegraphics[scale=0.4]{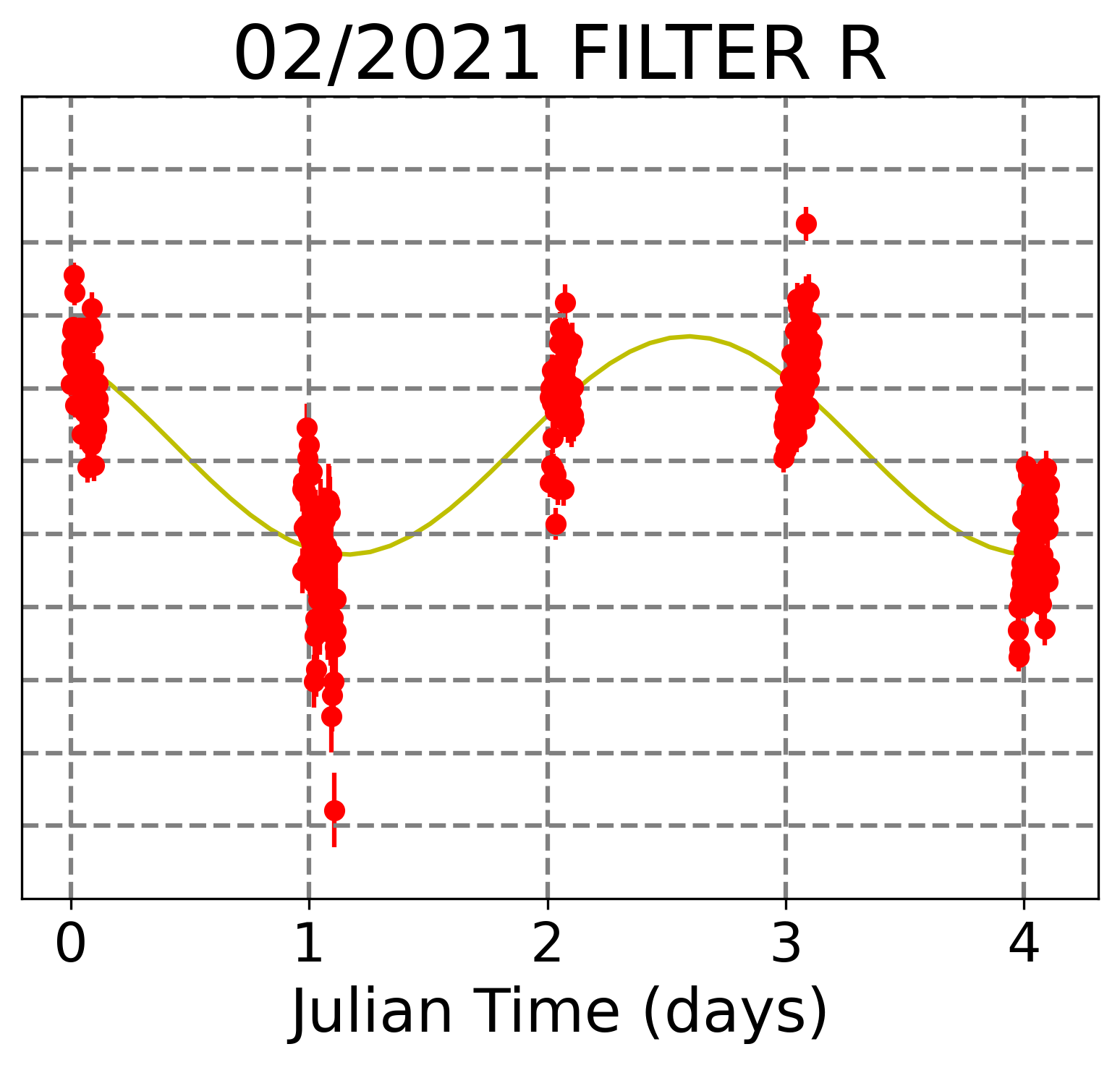}\\

            \includegraphics[scale=0.4]{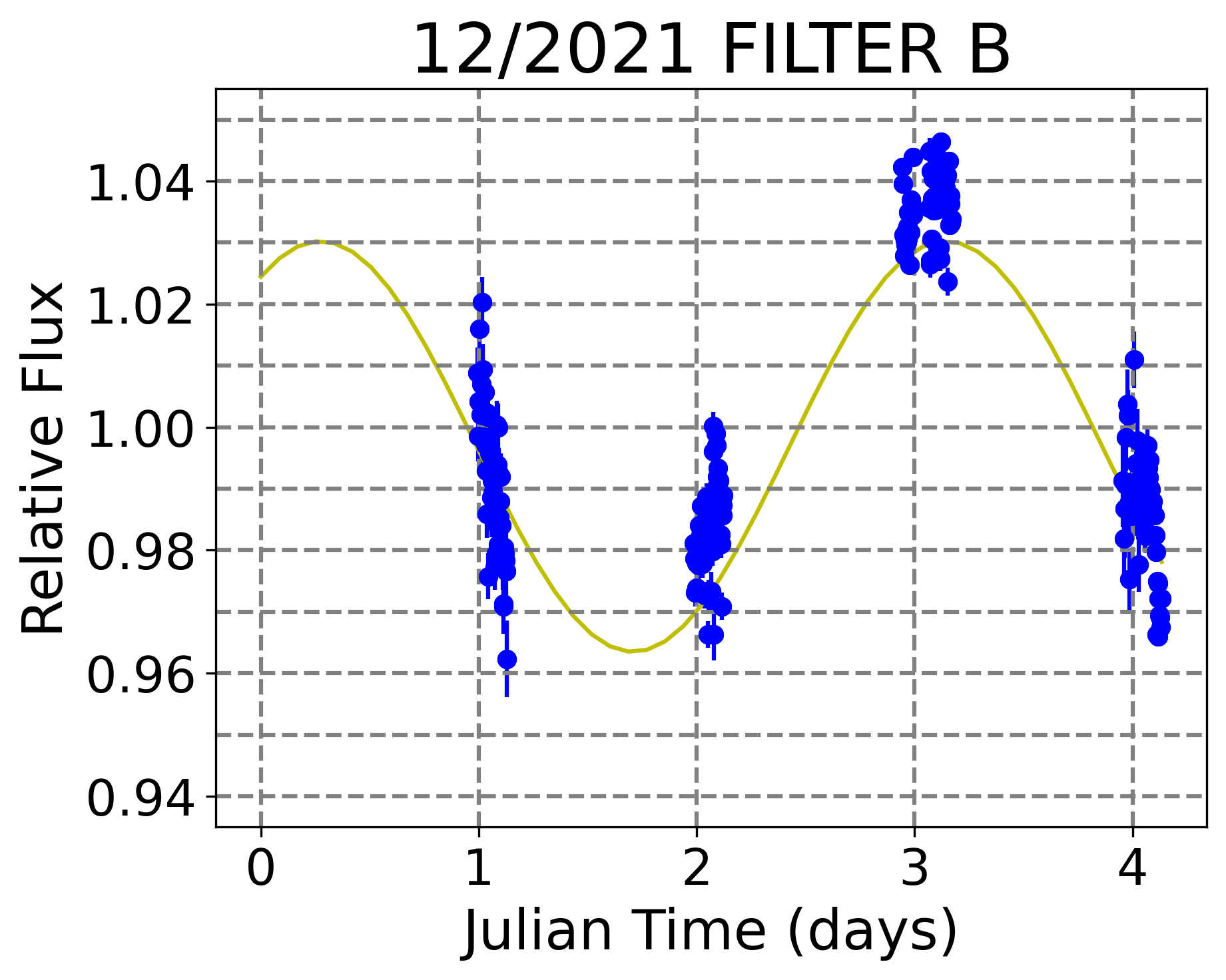}
            \includegraphics[scale=0.4]{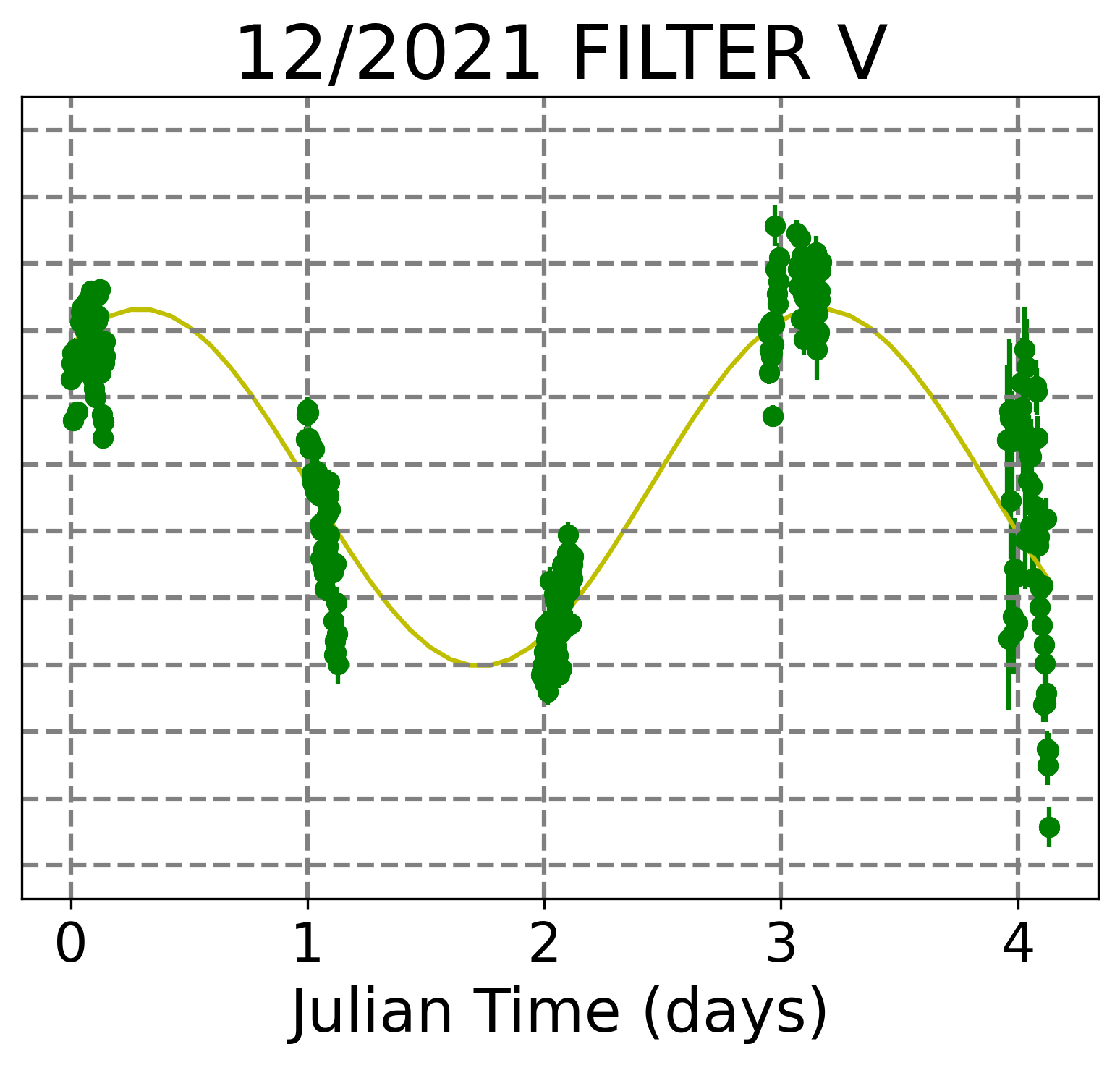}
            \includegraphics[scale=0.4]{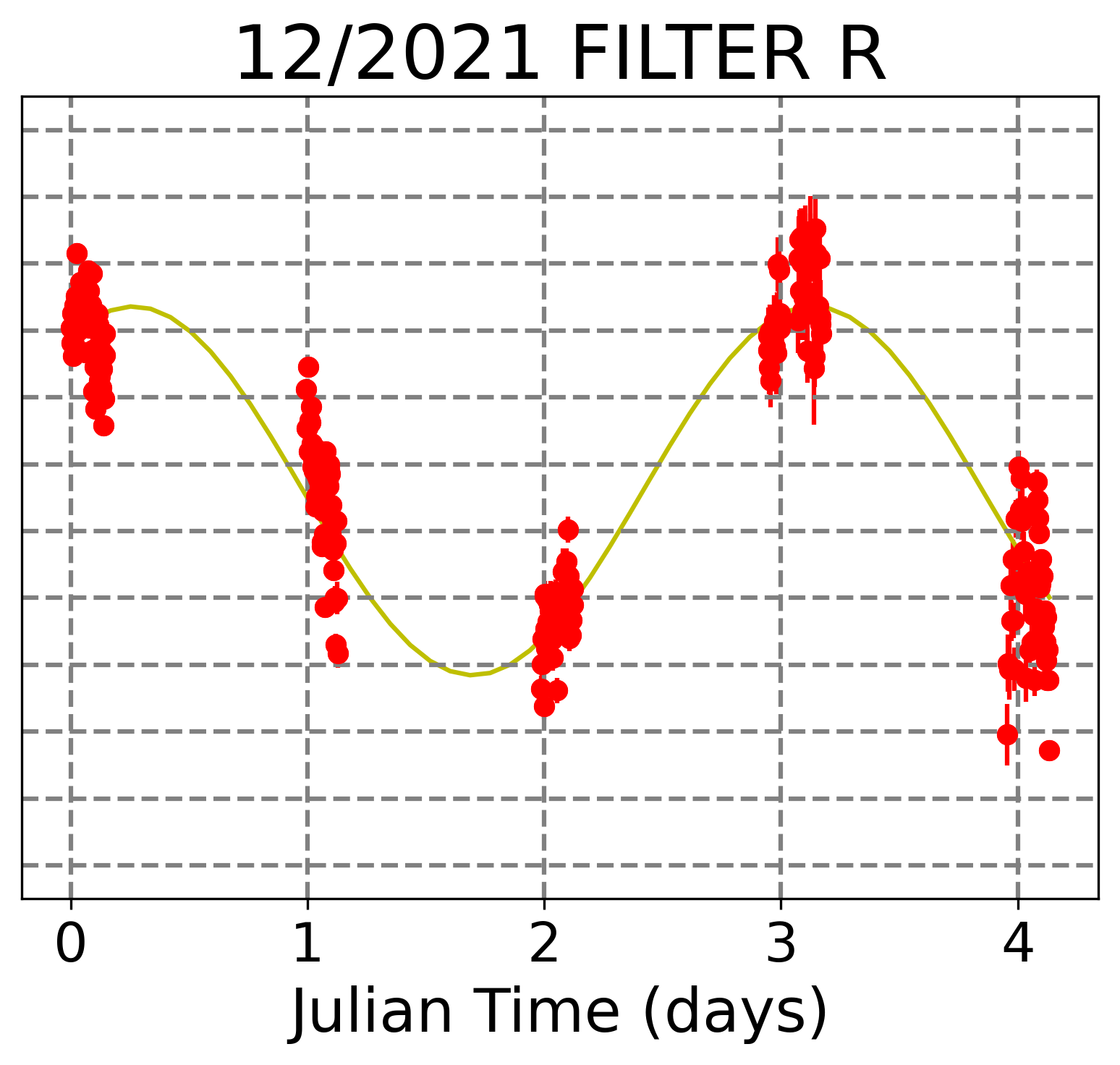}\\
            \includegraphics[scale=0.4]{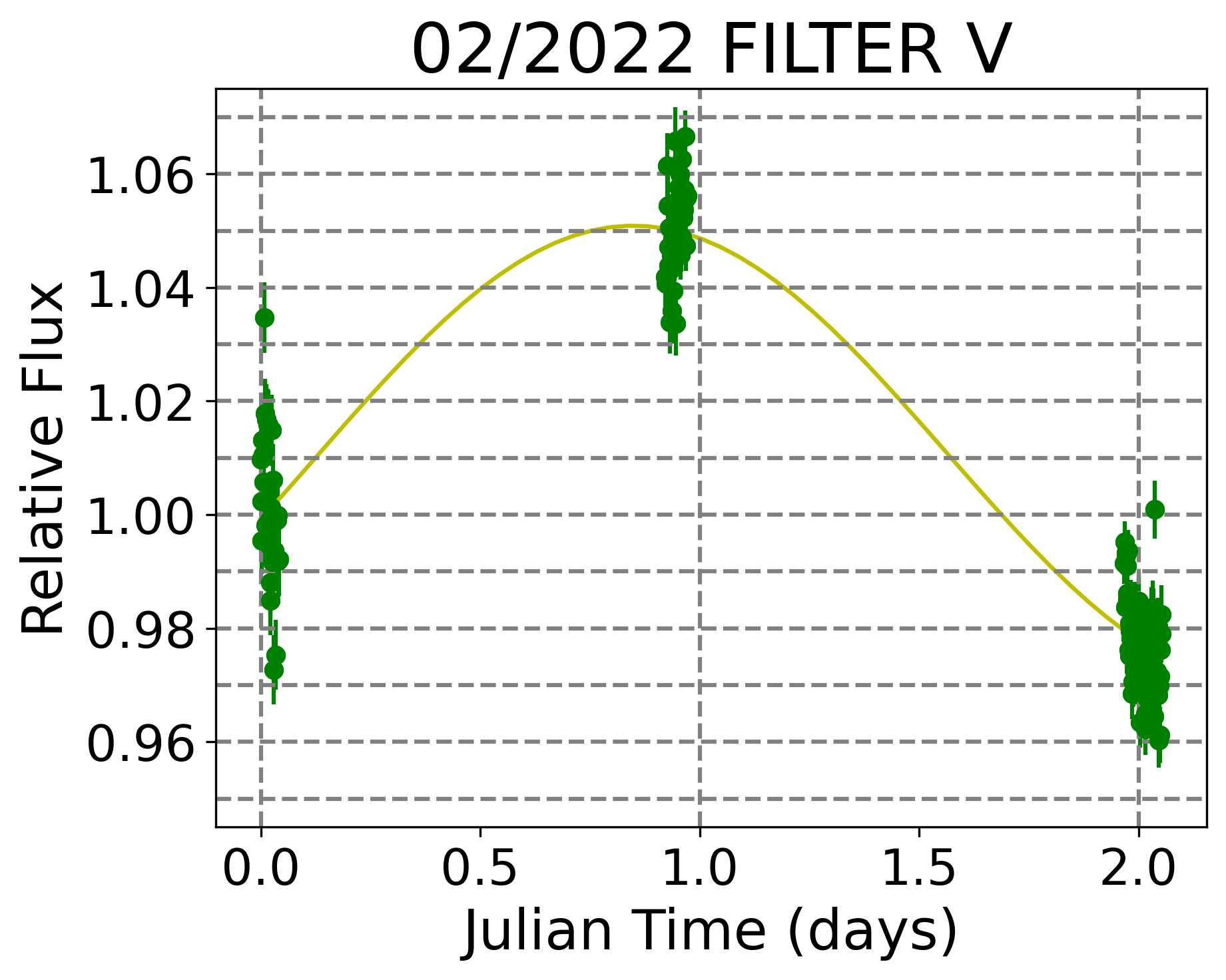}
            \includegraphics[scale=0.4]{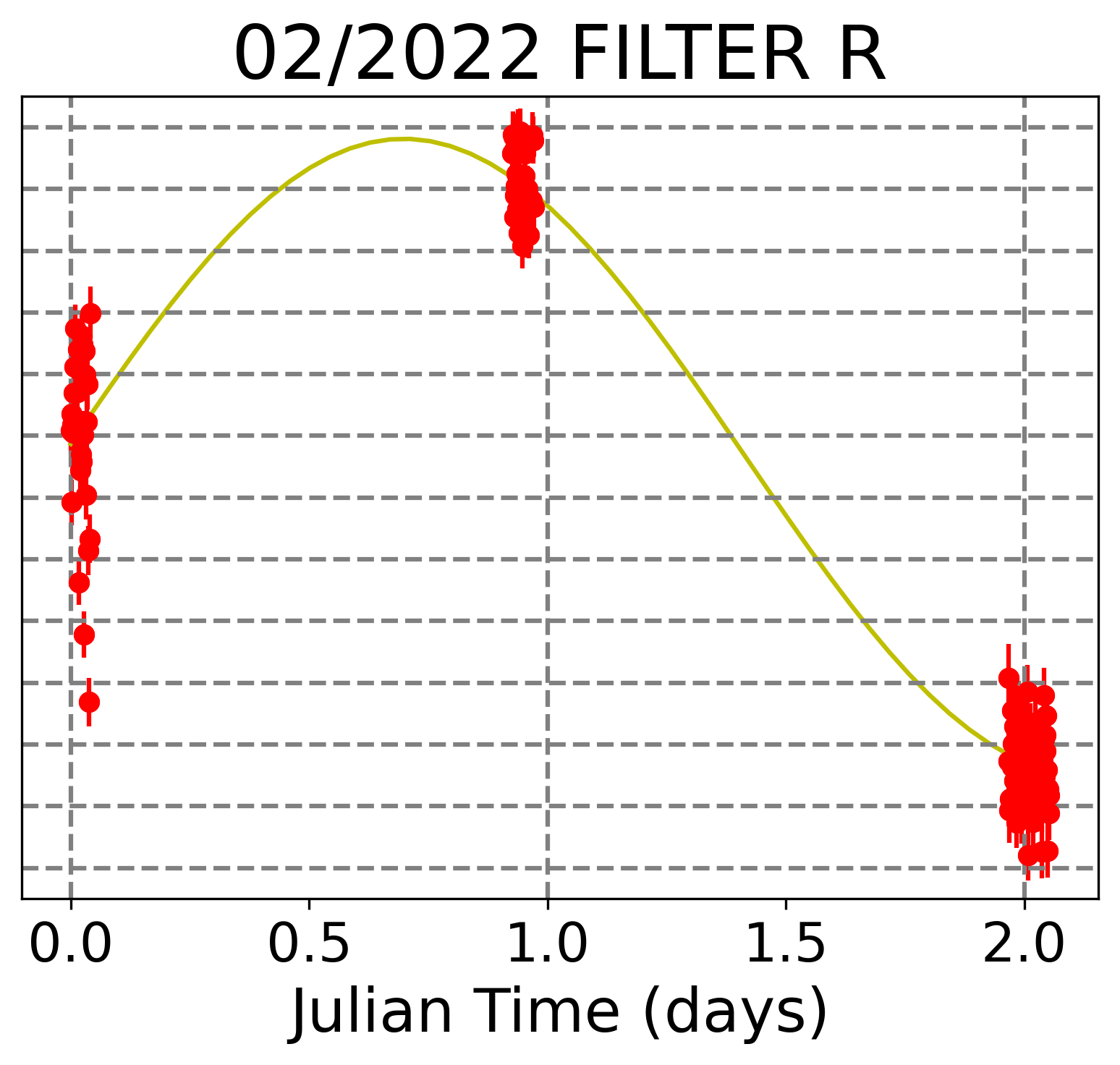}
            \includegraphics[scale=0.4]{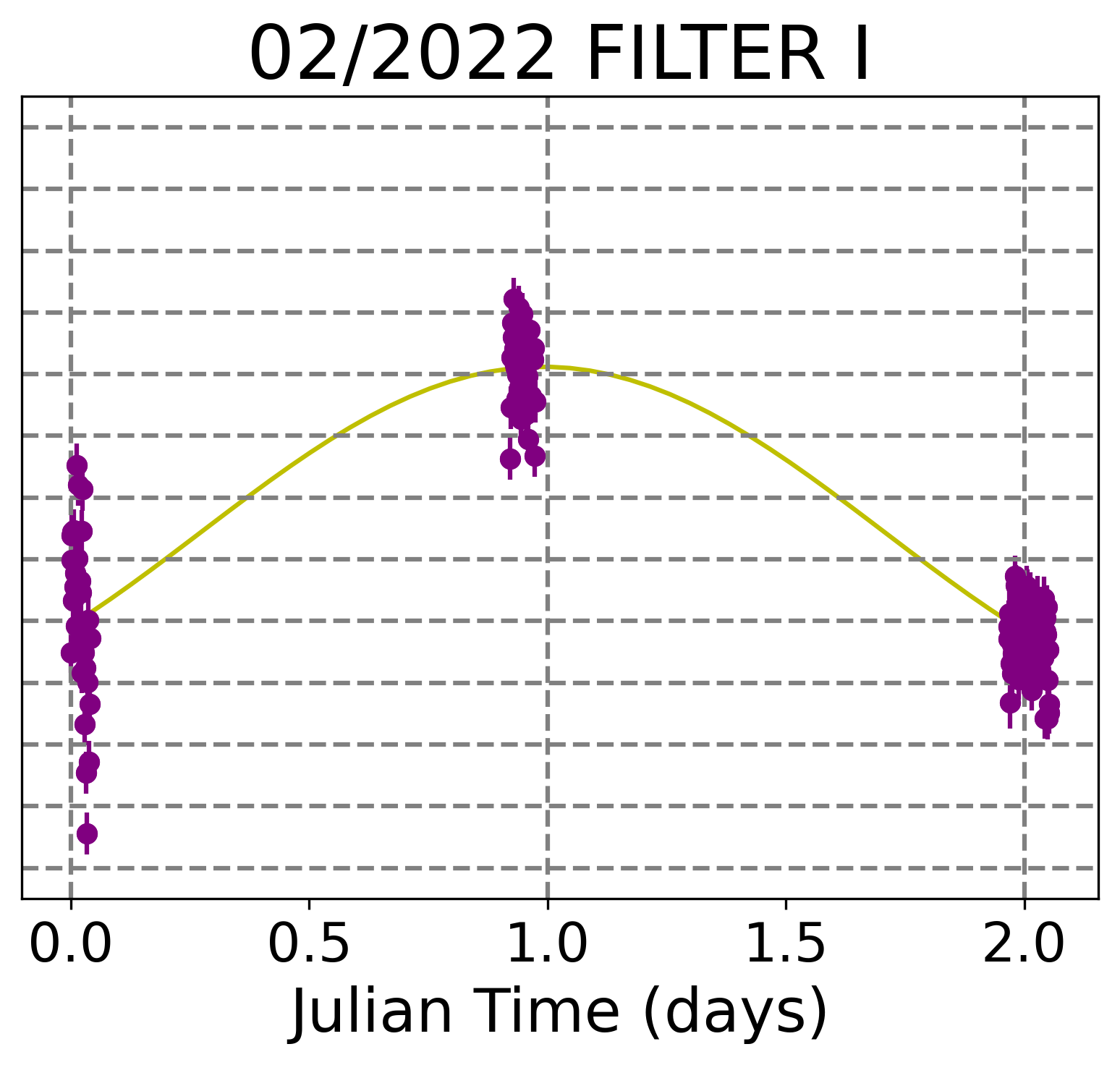}

        \caption{\label{DATA1} Data observed from OPC in B, V, R bands in February 2021 and December 2021 and in V, R and I bands in February 2022. The plotted line is an estimation of the flux variation of the star obtained by fitting the data by using a sinusoidal function with an offset. For the observations of February 2022 and for the first day of observation of December 2021, B band data were discarded because they had a signal-to-noise ratio (SNR) < 20.}
\end{figure*}

\begin{table}[!h]
\caption{Flux variations amplitudes measured analyzing V1298 Tau data in different photometric bands, obtained fitting the data with a sinusoidal function and an offset.}
\label{PERIODS}
\centering
\begin{tabular}{ c | c | c }
\hline
\hline
\rule{0pt}{4ex} Observing Run&Filter&Flux Variation\\
\hline
\rule{0pt}{4ex} 02/2021&B&$\mathrm{0.0515^{+0.0018}_{-0.0017}}$\\
\rule{0pt}{4ex} 02/2021&V&$\mathrm{0.0425^{+0.0013}_{-0.0013}}$\\
\rule{0pt}{4ex} 02/2021&R&$\mathrm{0.0356^{+0.0013}_{-0.0012}}$\\
\rule{0pt}{4ex} 12/2021&B&$\mathrm{ 0.084^{+0.002}_{-0.002}}$\\
\rule{0pt}{4ex} 12/2021&V&$\mathrm{0.066^{+0.002}_{-0.002}}$\\
\rule{0pt}{4ex} 12/2021&R&$\mathrm{0.0660^{+0.0018}_{-0.0018}}$\\
\rule{0pt}{4ex} 02/2022&V&$\mathrm{0.096^{+0.002}_{-0.003}}$ \\
\rule{0pt}{4ex} 02/2022&R&$\mathrm{0.1194^{+0.0019}_{-0.0019}}$ \\
\rule{0pt}{4ex} 02/2022&I&$\mathrm{0.061^{+0.002}_{-0.002}}$ \\
\hline
\end{tabular}
\end{table}
\begin{table}[!t]
\caption{Priors on spot parameters of MultiNest for our retrieval procedure to model V1298 activity. Spot temperature's prior is set with a maximum temperature exceeding the unspotted surface one to include the possibility of retrieving a facula instead of a spot.}
\label{PRIORS}
\centering
\begin{tabular}{ c | c }
\hline
\hline
\rule{0pt}{3ex}\textbf{Parameter} & \textbf{Priors}\\
\hline
\rule{0pt}{4ex} Jitter &1e-4,1e-1\\
\rule{0pt}{4ex} Scale Factor &0.5-5\\
\rule{0pt}{4ex} Temperature &3000-5500 K\\
\rule{0pt}{4ex} Latitude &0-90°\\
\rule{0pt}{4ex} Longitude &-180-180°\\
\rule{0pt}{4ex} Radius &0-1.0 $\mathrm{R_{\star}}$ \\
\hline
\end{tabular}
\end{table}
We calibrated and analyzed the images using AstroImageJ \citep[a software developed by TESS collaboration,][]{2017AJ....153...77C} and we applied the so-called differential photometry method: we did not measure the absolute photometry of the target star, but its variability with respect to some carefully chosen stable check stars in the same field of view of V1298 Tau, used as reference. The choice of these check stars required a delicate compromise between the necessity of the highest number of them and the requirement of stability and comparable magnitude with respect to the target. In the FoV of the telescope (20' x 20') around the star there are only a few stars and most of them were saturated, very faint or variable, so in the end we found only 6 stable reference stars. It should be noted that we used the same reference stars for all the datasets.\\
Using this procedure we obtained the light-curve of V1298 Tau in the selected photometric bands shown in Figure \ref{DATA1}. 
To retrieve the rotational period of the star, we employed a sinusoidal function and an offset to fit the data. This function feeds the nested sampling algorithm \textit{Multinest v3.10} \citep{2009MNRAS.398.1601F} to derive parameters best-fit values and errors throughout the python package \textit{PyMultiNest v2.12} \citep{2014A&A...564A.125B} and with the following likelihood:

\begin{equation}
\begin{split}
        \ln{p({y_n},{t_n},{\sigma},\theta)} =& -1/2\sum_n \left[ (y_n-F(t_n,\theta))^2/\sigma^2 + \ln2 \pi \sigma^2 \right]
\end{split}
\end{equation}

where $y_n$ and $t_n$ are, respectively, data fluxes and times, $\sigma=\sqrt{\sigma_n^2+\sigma_j^2}$ with  $\sigma_n$ as data errors and $\sigma_j$ as white noise jitter term, introduced to account for unknown source of errors. The algorithm was set to use 1000 live points, leaving all other parameters to their default values.

This analysis allowed to estimate the flux variations (see Table \ref{PERIODS}) of the star and verify the coherence in each observation.

Therefore, from Figures \ref{DATA1} we observed variability in the amplitude of the sinusoidal fit, which on average tends to decrease at longer wavelengths, consistent with the assumption of a spot-dominated star.  The results of this analysis are summarized in Table ~\ref{PERIODS}.

\begin{figure*}[!h]
		\centering
		\includegraphics[scale=0.3]{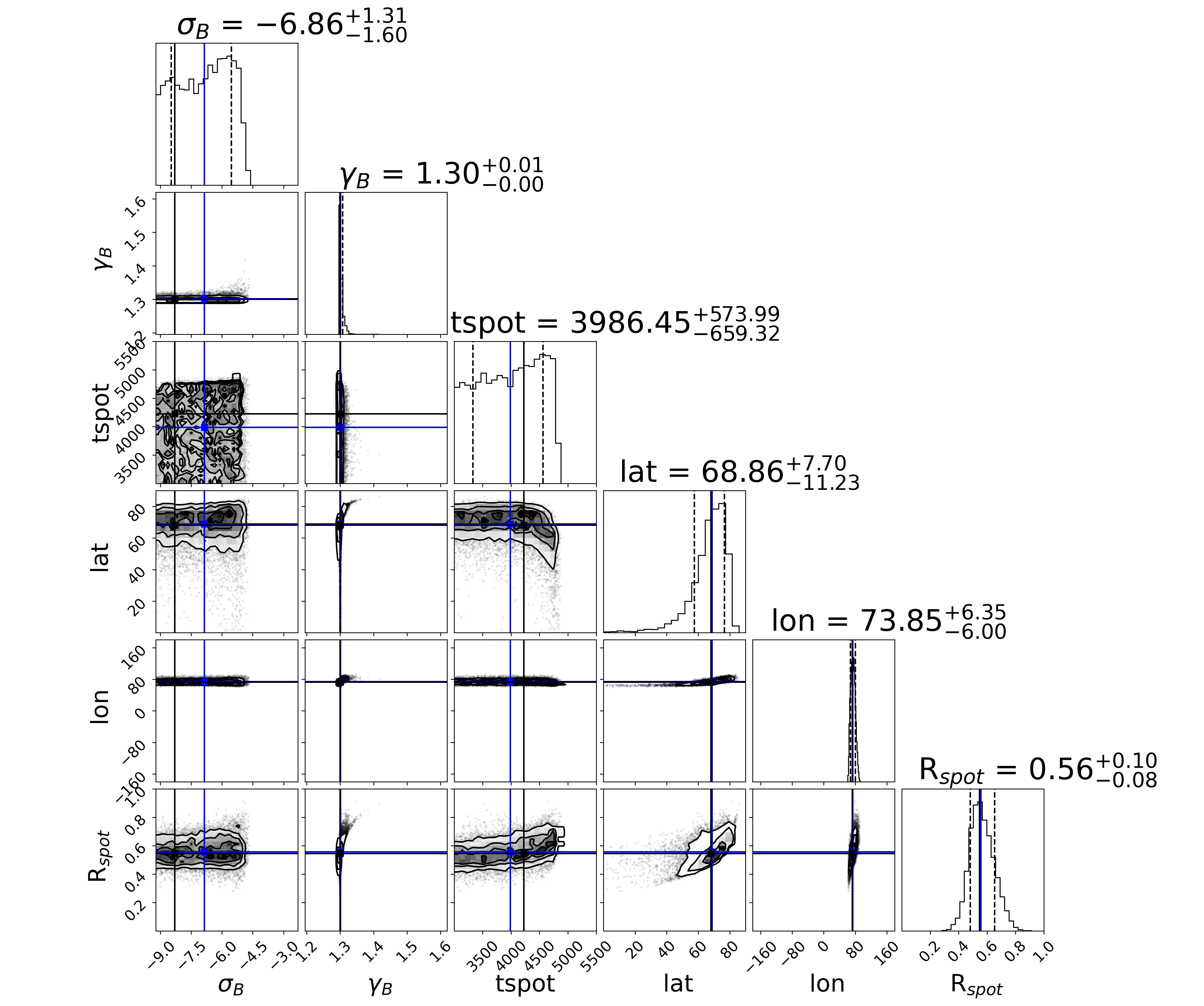}
            \includegraphics[scale=0.3]{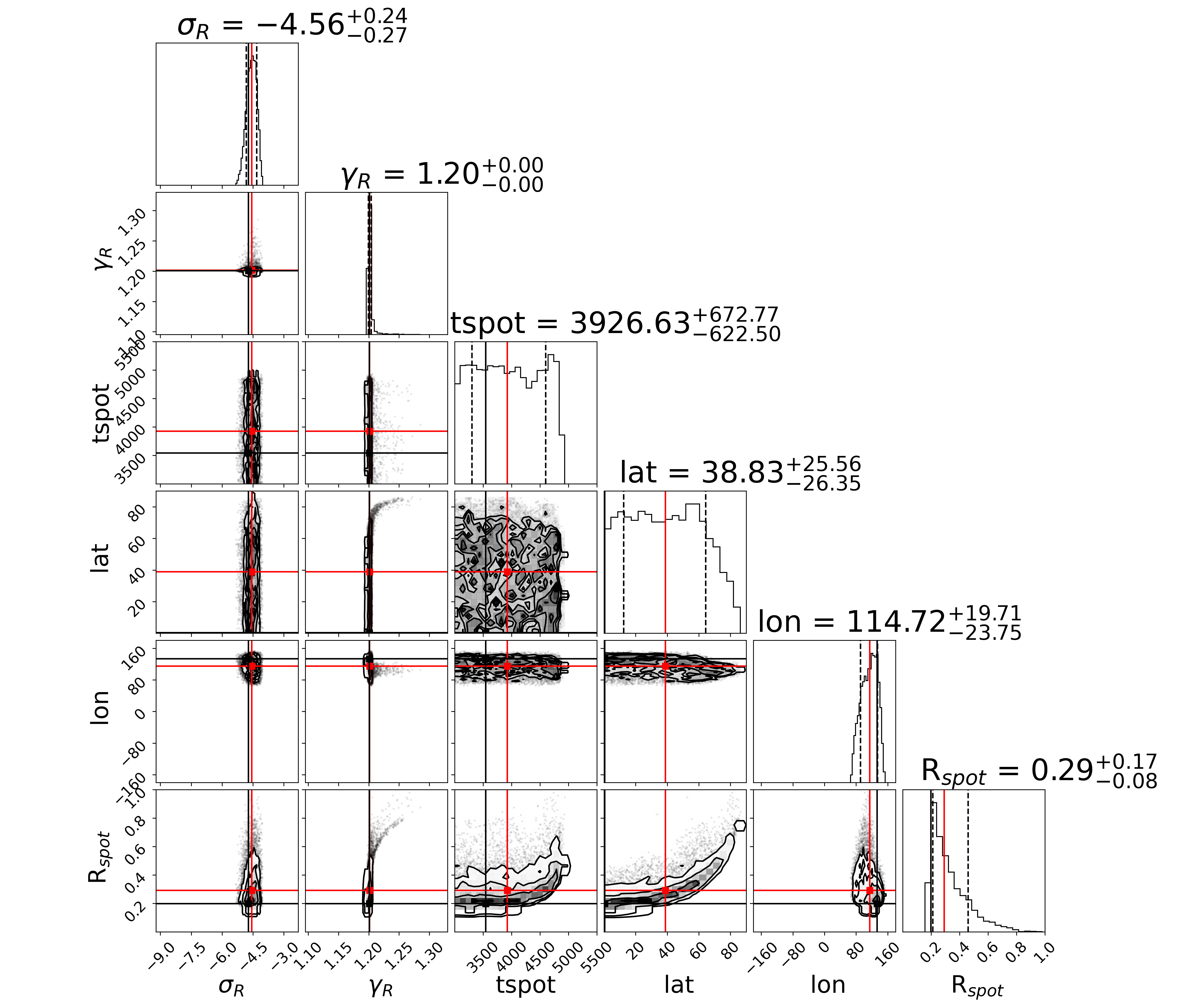}
		\caption{Comparison of the retrieval posteriors of spots for V1298 Tau, obtained by OPC observations on 23-24-25/02/2021 through a separate analysis for data in B (top) and R (bottom) photometric bands. Blue lines (for right image) and red lines (for left image) mark median values, while black lines mark maximum probability (MAP) values for both images.}
            \label{Old}
\end{figure*}%

\begin{figure*}[!t]
		\centering
		\includegraphics[width=\linewidth]{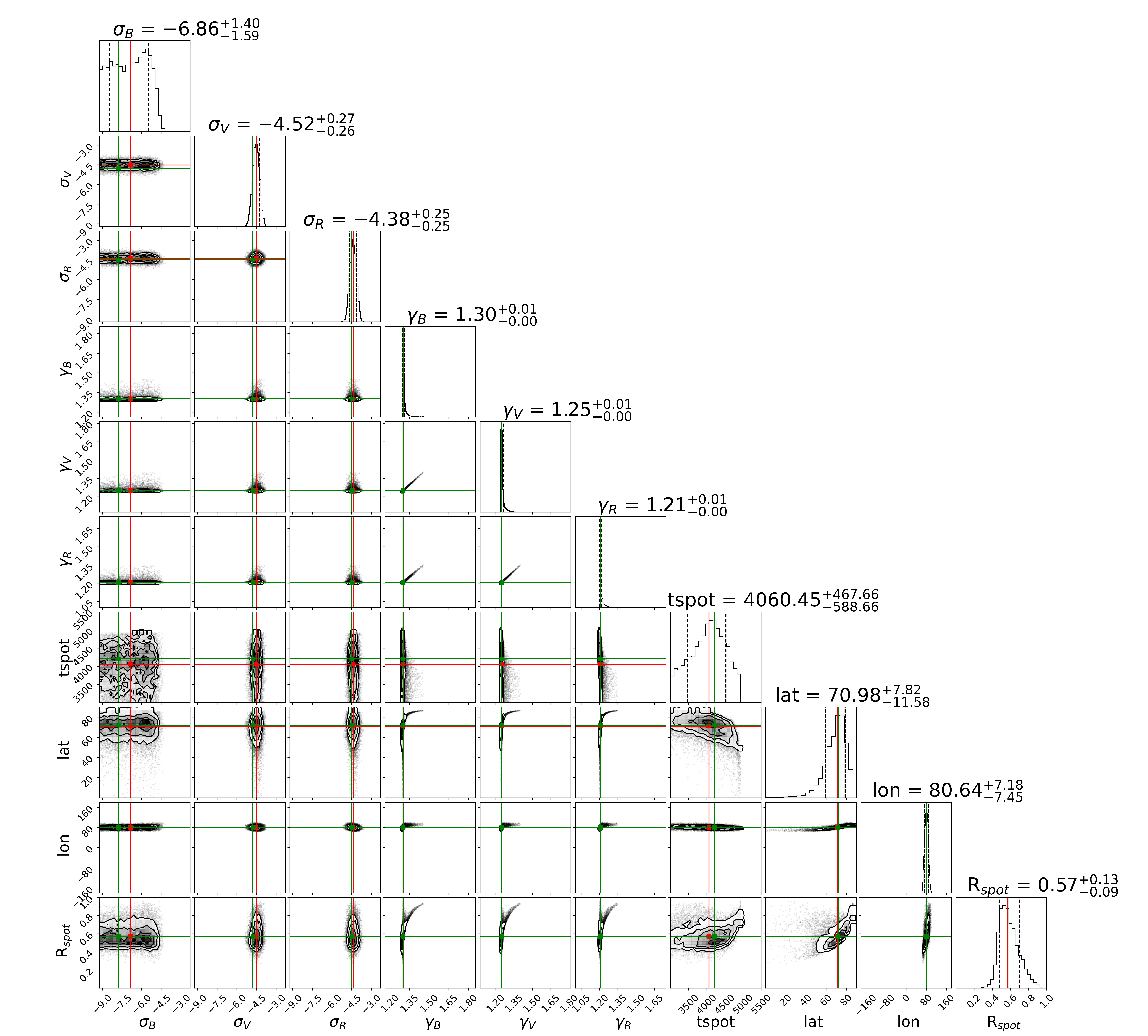}
            \caption{Corner plot obtained by the retrieval procedure for the multiband photometric data acquired by OPC in the first run of observation, specifically on February 23-24-25th, 2021. $\mathrm{\sigma_{i}}$ are the jitter parameters and $\mathrm{\gamma_{i}}$ are the offsets of the forward model used in the retrieval for each band observed (B, V and R). Red lines mark median values, while green lines mark MAP values. }
            \label{CORN1}
\end{figure*}

\begin{figure*}[!t]
		\centering
		\includegraphics[width=\linewidth]{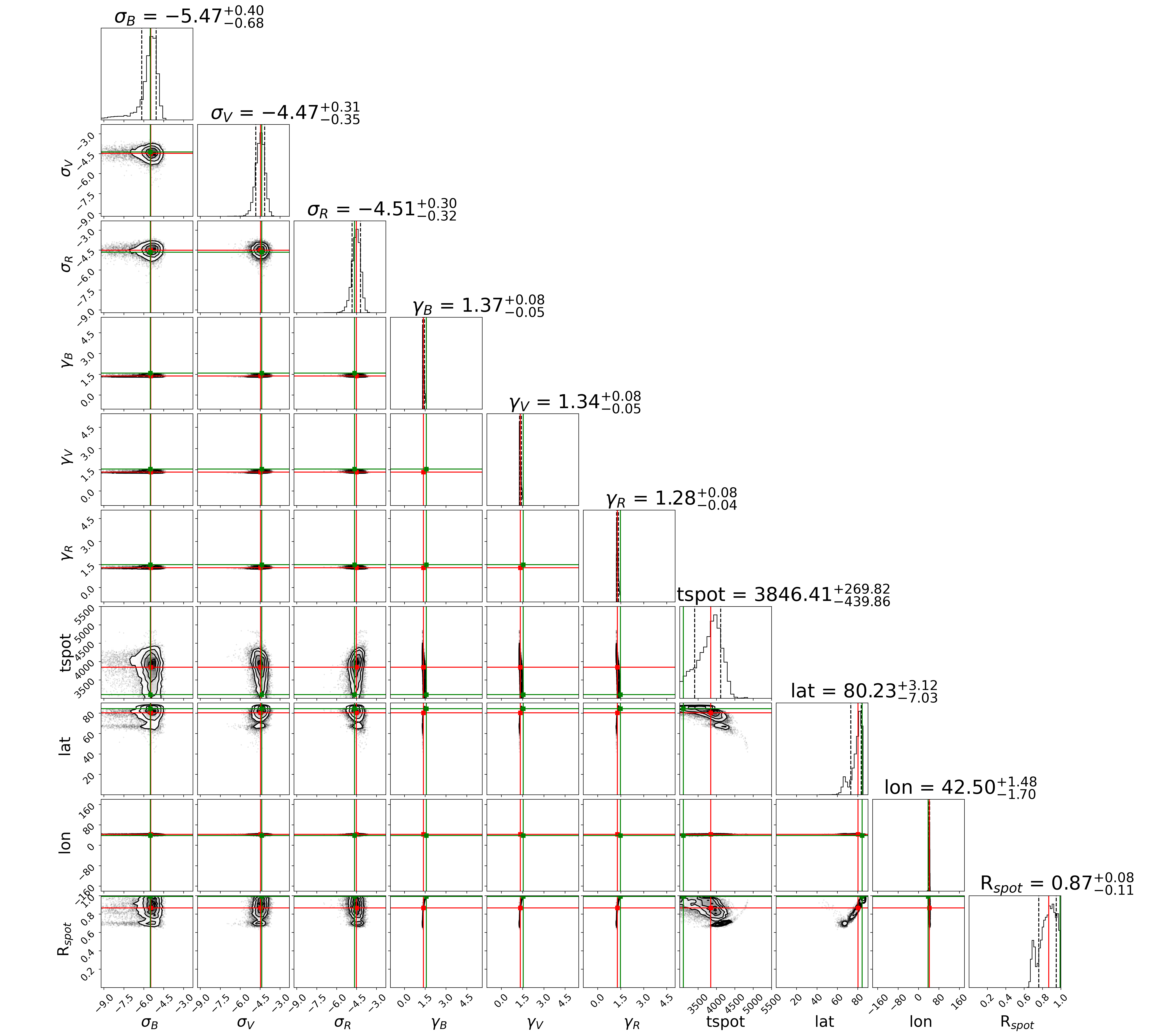}
            \caption{Same as for Figure \ref{CORN1}, but in the case of B, V and R bands on December 13-14-15th, 2021.}
            \label{CORN2}
\end{figure*}
\begin{figure*}[!t]
		\centering
		\includegraphics[width=\linewidth]{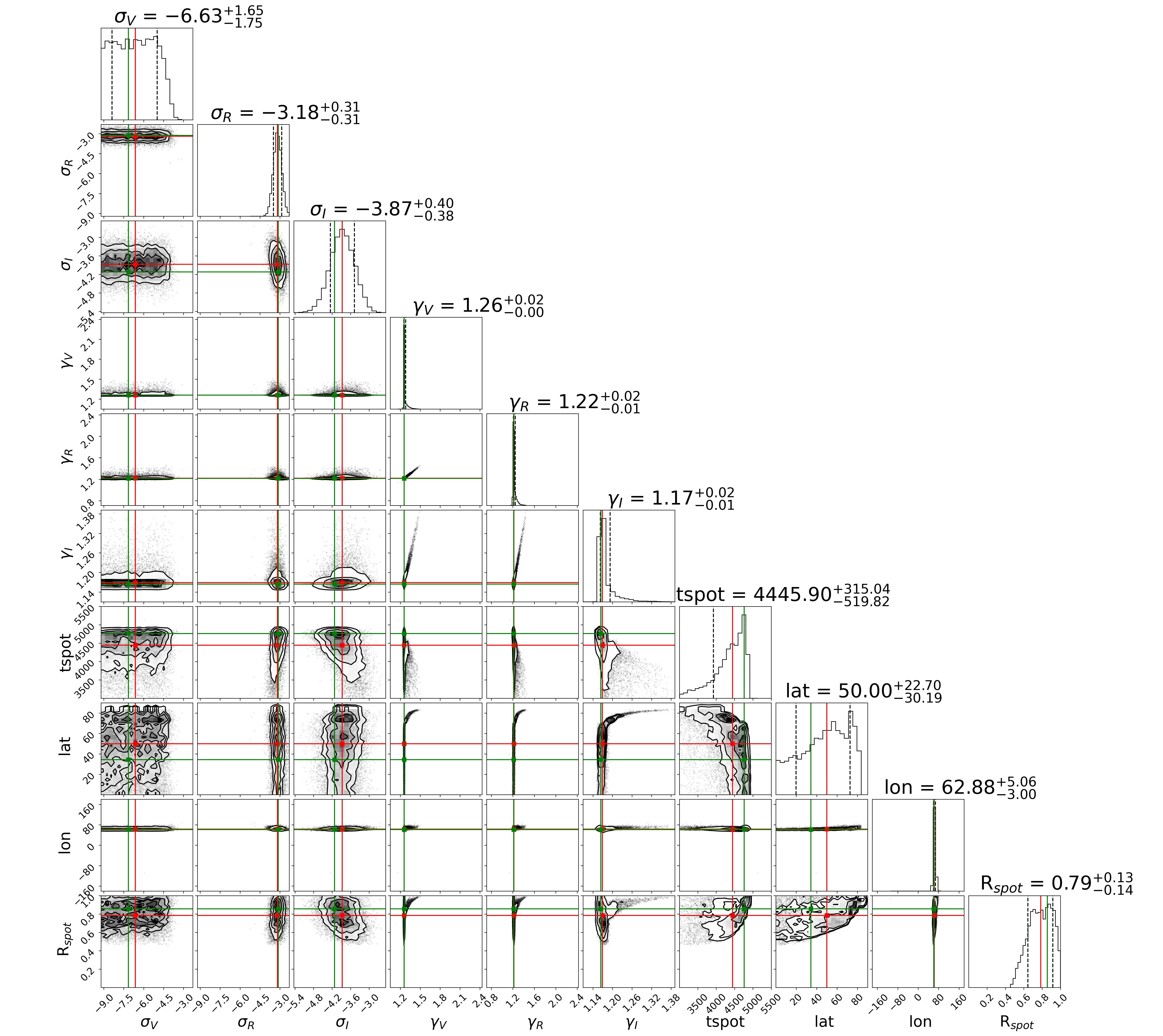}
            \caption{Same as for Figure \ref{CORN1}, but in the case of V, R and I bands on February 21-22-23rd, 2022.}
            \label{CORN3}
\end{figure*}

\begin{table}[!h]
\caption{Spot temperature range estimates for each independent set of observations in the hypothesis of one spot. For the set marked with "*" we used only V and R data because B data had a low SNR, while for the same reason we used V, R and I data for the last set (marked with "**"). }
\label{Temperatures}
\centering
\begin{tabular}{ c | c | c }
\hline
\hline
\rule{0pt}{3ex}\textbf{Dates} & \textbf{Bands} & \textbf{Tspot (K)}\\
\hline
\rule{0pt}{4ex} 21-22-23/02/2021 &B-R-V&$4267^{+273}_{-416}$ \\
\rule{0pt}{4ex} 22-23-24/02/2021 &B-R-V&$4433^{+247}_{-362}$ \\
\rule{0pt}{4ex} 23-24-25/02/2021 &B-R-V&$4060^{+468}_{-589}$ \\
\rule{0pt}{4ex} 11-12-13/12/$2021^{*}$&R-V&$3618^{+468}_{-420}$ \\
\rule{0pt}{4ex} 12-13-14/12/2021&B-R-V&$4176^{+250}_{-308}$ \\
\rule{0pt}{4ex} 13-14-15/12/2021&B-R-V&$3846^{+270}_{-440}$ \\
\rule{0pt}{4ex} 21-22-23/02/$2022^{**}$ &R-V-I&$4446^{+315}_{-520}$ \\
\rule{0pt}{1ex} &&\\
\hline
\end{tabular}
\end{table}	

\begin{table}[!h]
\caption{Spot temperature range estimates for each observing run in the 1 spot hypothesis. For observing runs longer than 3 days, we took the common range of the various 3 days-sequences of that run.}
\label{TEMPSET}
\centering
\begin{tabular}{ c | c }
\hline
\hline
\rule{0pt}{3ex}\textbf{Periods} & \textbf{Tspot (K)}\\
\hline
\rule{0pt}{4ex} 21-25/02/2021 &4071-4528 K \\
\rule{0pt}{4ex} 11-15/12/2021 &3868-4086 K \\
\rule{0pt}{4ex} 21-23/02/2022 &3926-4761 K \\
\hline
\end{tabular}
\end{table}


\begin{figure}
		\centering
		\includegraphics[width=\columnwidth]{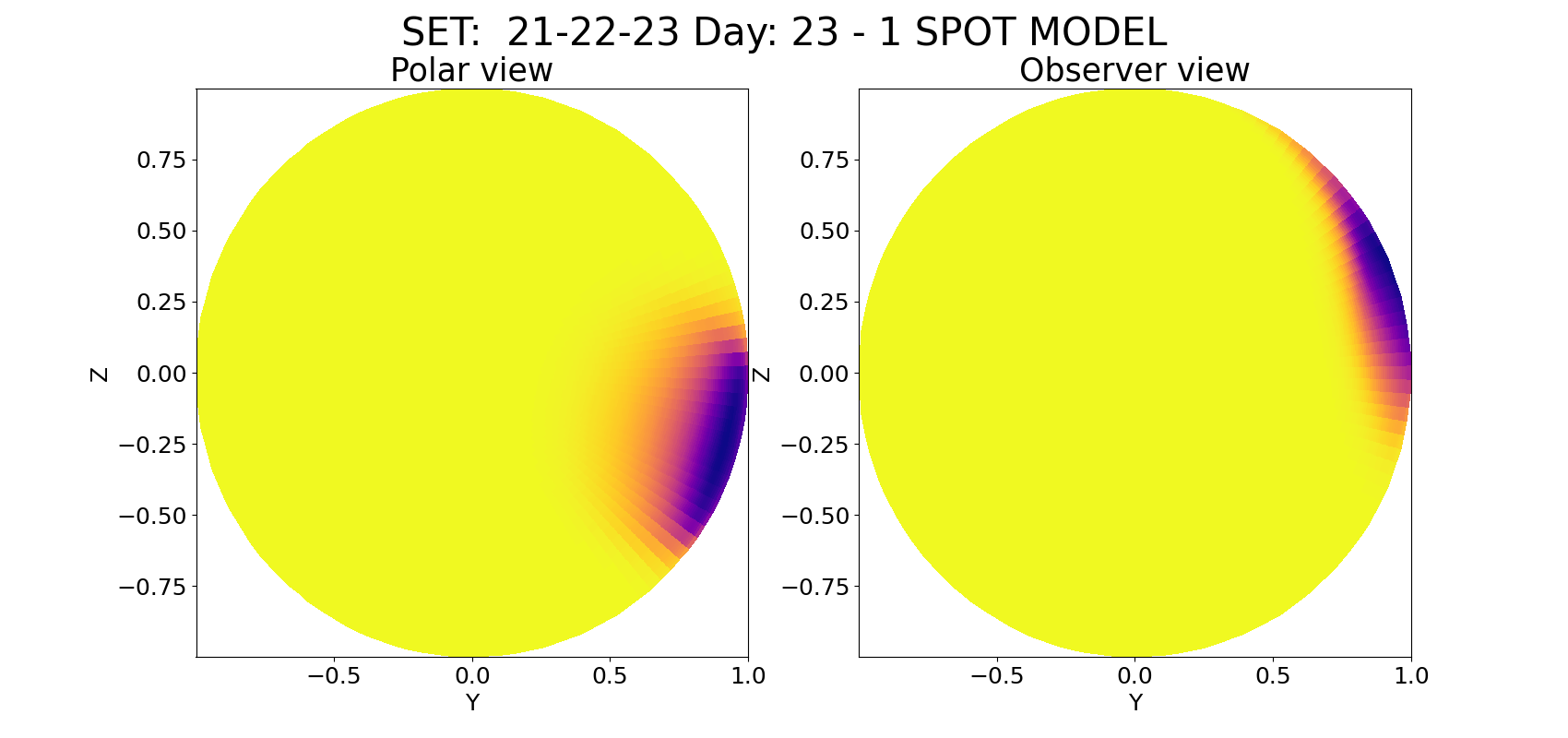}
            \includegraphics[width=\columnwidth]{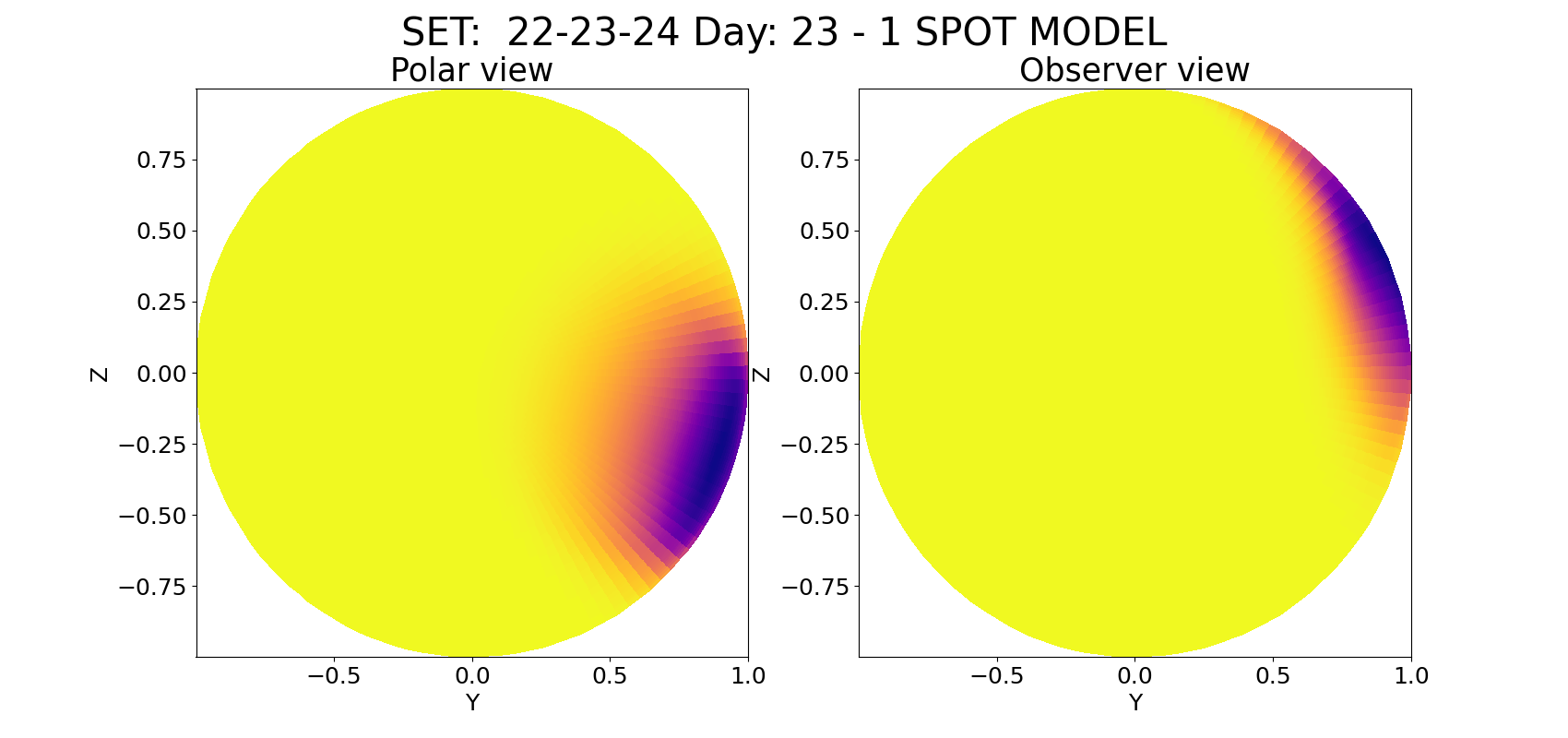}
		\includegraphics[width=\columnwidth]{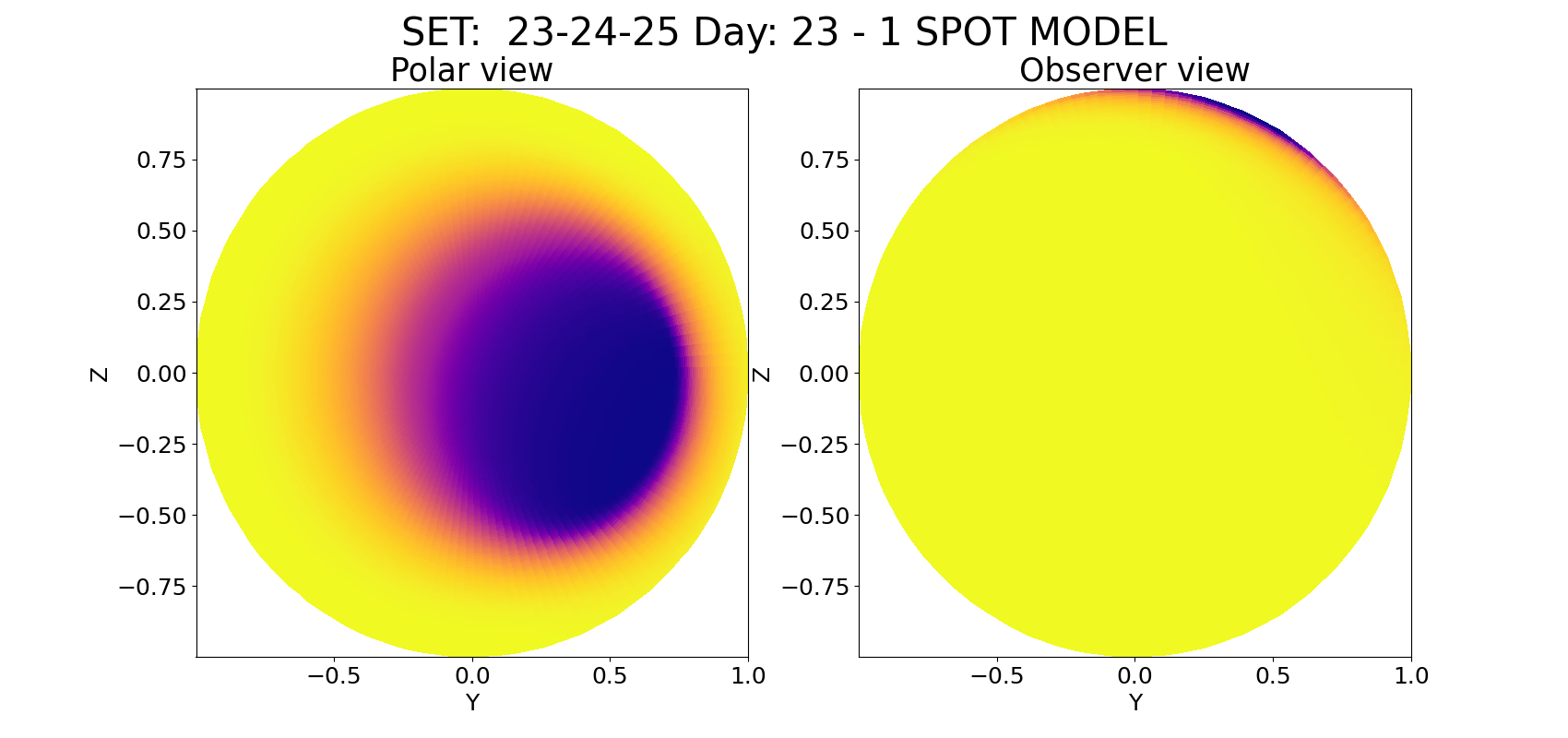}
		\caption{\label{Geometry} V1298 Tau spots positions of the 3 observing sets of February 2021, given by the distribution of solutions of MultiNest retrieval. Each image shows the star as seen on the same day, February 23th 2021. The spots are shown both from a polar view (on the right) and from the equator, i.e. from Earth (on the left).} 
\end{figure}%

\begin{figure}[!t]
		\centering
		\includegraphics[width=\columnwidth]{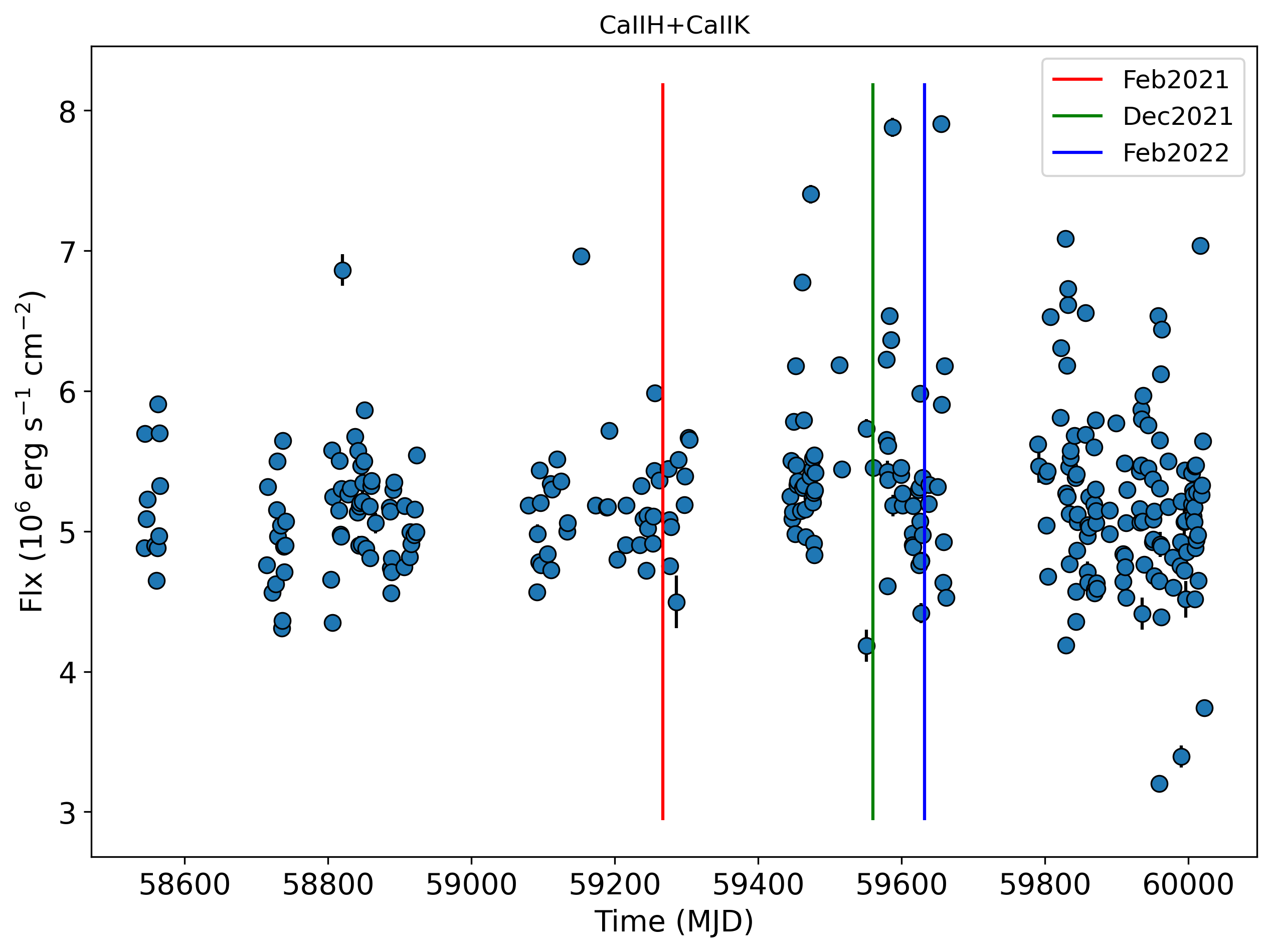}
		\caption{\label{CaHK} CaIIH + CaIIK values obtained by HARPS-N for V1298 Tau. The coloured lines correspond to our periods of observations: February 2021, December 2021 and February 2022.}
\end{figure}
\begin{figure}[!t]
		\centering
		\includegraphics[width=\columnwidth]{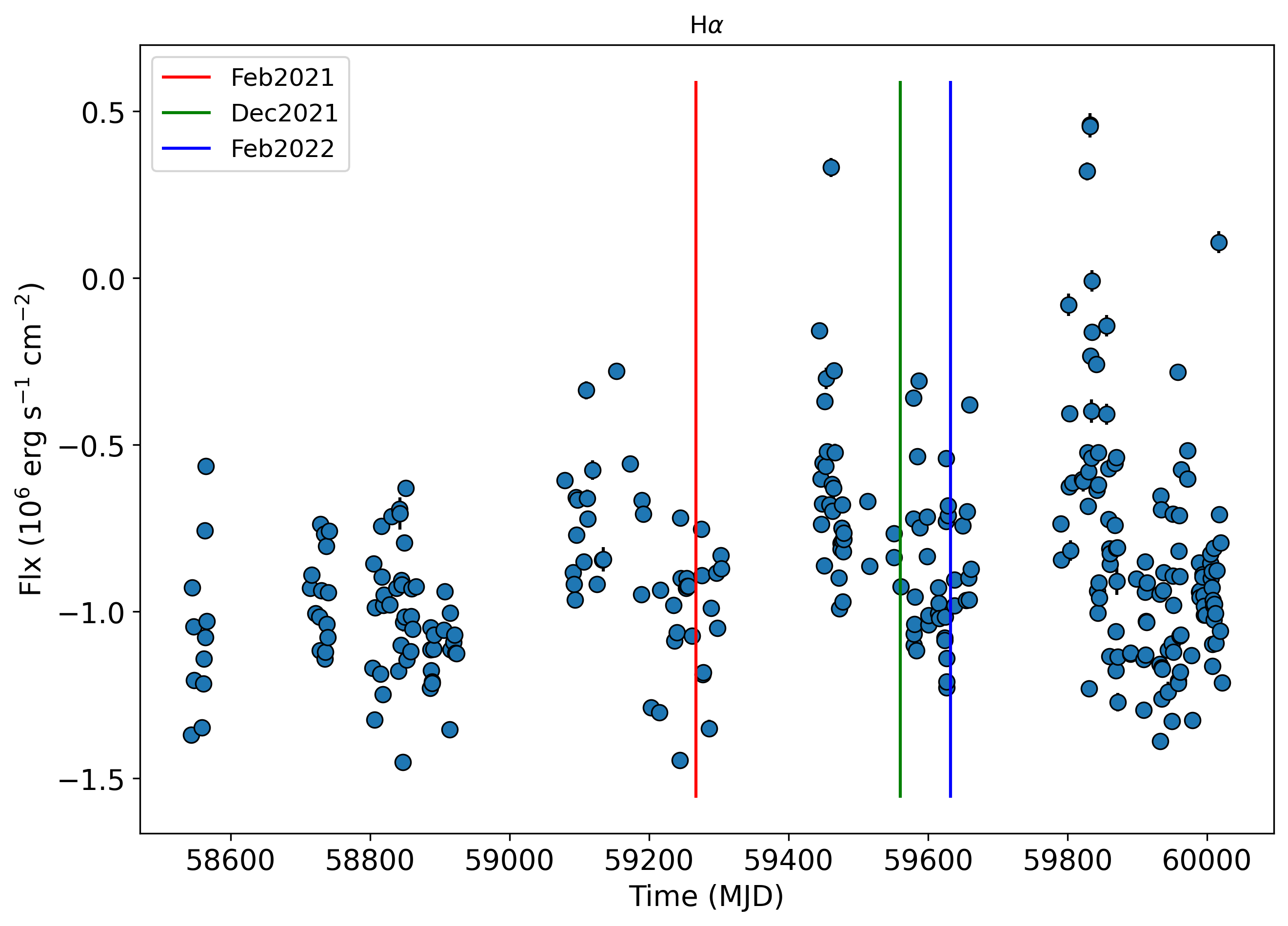}
		\caption{\label{Halpha} H alpha values obtained by HARPS-N for V1298 Tau. The coloured lines correspond to our periods of observations: February 2021, December 2021 and February 2022.}
\end{figure}
\begin{figure}[!t]
		\centering
		\includegraphics[width=\columnwidth]{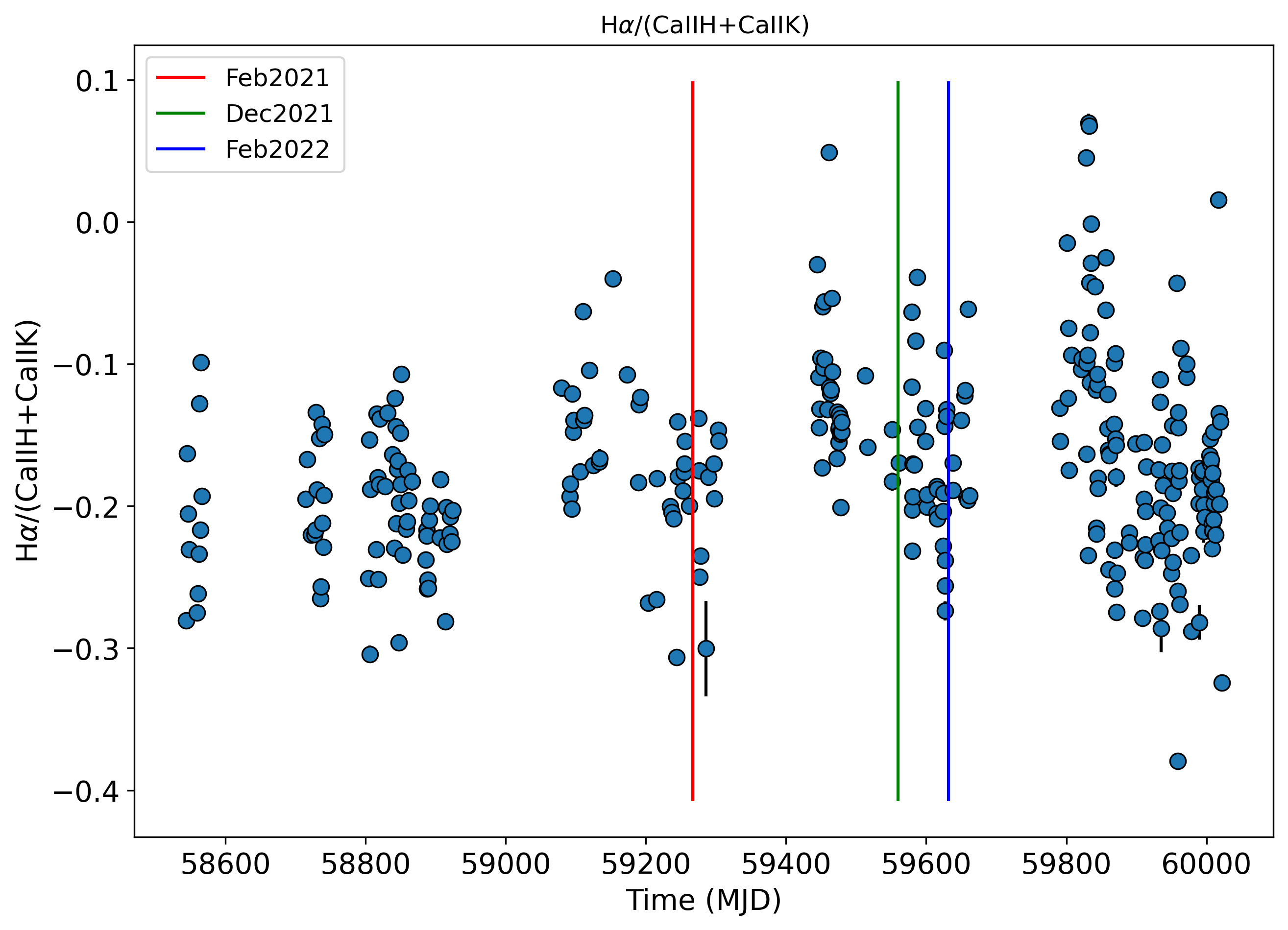}
		\caption{\label{HaCa} Ratio of $\mathrm{H\alpha/(CaII H +CaII K)}$ values obtained by HARPS-N spectrometer for V1298 Tau. The coloured lines correspond to our periods of observations: February 2021, December 2021 and February 2022.}
\end{figure}

\begin{figure*}
    \centering
    \includegraphics[width=\linewidth]{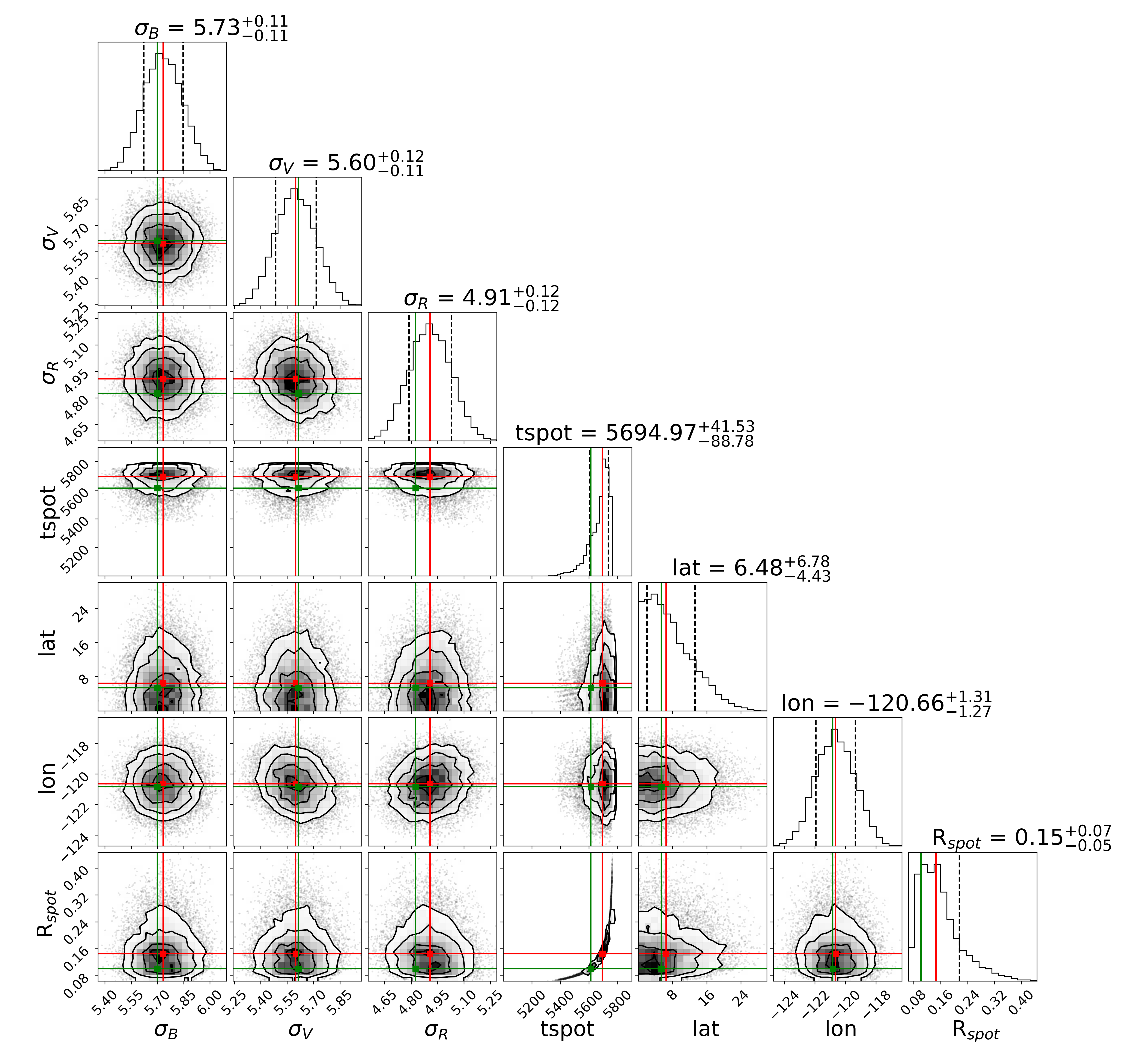}
    \caption{Posteriors of the retrieval procedure for the analysis of the active region properties on the Sun during the first half of the rotational period starting on 12/30/2013. Green lines mark MAP values, while red lines mark median values.}
    \label{CORNERSUN}
\end{figure*}

\begin{figure*}
    \centering
     \includegraphics[scale=0.68]{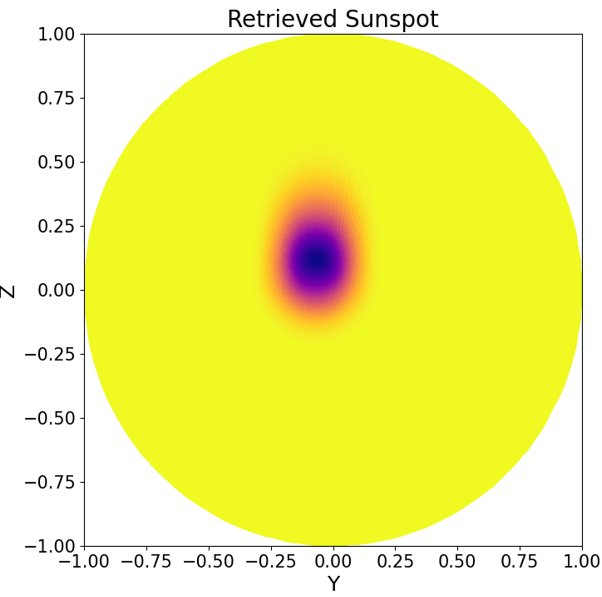}
    \includegraphics[scale=0.45]{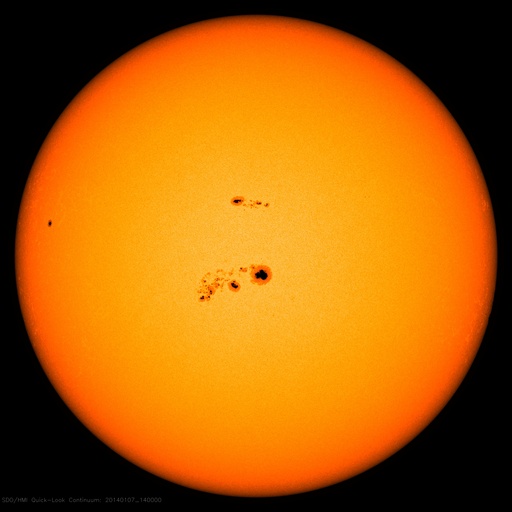}\\
    \includegraphics[scale= 0.8]{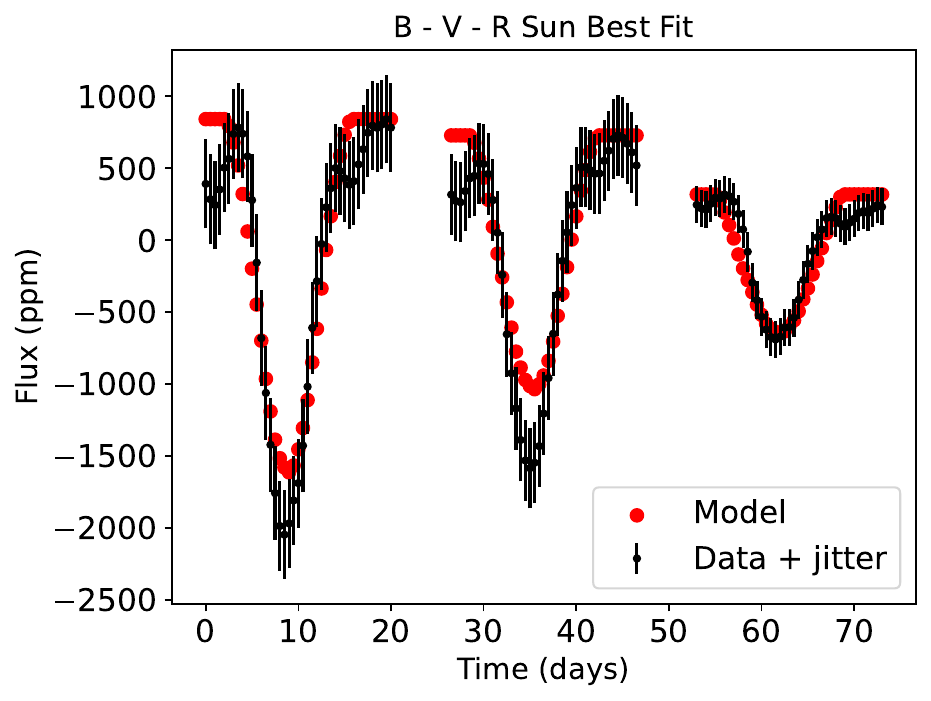}
    \caption{Upper-left: simulated sunspot according to our model results, shown in Figure \ref{CORNERSUN}. Upper-right: a real image of the Sun during the period studied, from HMI.
    Bottom: best fit of our model compared to the observed data for the Sun in all the three bands, from left to right, B, V and R filter, shifted in time.}
    \label{SUN}
\end{figure*}

\section{Forward Model and Retrieval}
We modelled the spot properties (latitude, longitude, radius and the temperature of the spots) following the approach of \cite{2021MNRAS.507.6118C,2021MNRAS.501.1733C} and we imposed some assumptions: 
\begin{itemize}
	\setlength\itemsep{0em} 
	\item  Spots corotating with the stellar surface
	\item  Not evolving spots during a rotational period of the star
	\item  Approximation of circular spots projected on the stellar surface
	\item  Same temperature for all the spots     
\end{itemize} 
With these hypotheses, we aimed to analyze flux variations of the star as generated by the luminosity contrast between the spots and the photosphere. 
Since the star has an inclination of nearly 90°, as we can see by combining the v sin i value of the star with its rotational period (see table \ref{V1298_properties}), we had a degeneration in latitude because we could not distinguish the hemisphere in which the specific spot was located. We removed instead the degeneracy in longitude for the spots thanks to the rotation of the star, so we could retrieve their longitude from the time variations of the stellar flux.
We chose to analyze datasets of 3 consecutive days (about a rotational period of the star) in order to avoid relevant contributions from spots evolution.
In order to build our forward model for the retrieval procedure, we divided the star into concentric rings, with a finer division near the limb of the stellar disk to better account for the effect of limb darkening.
We also took into account the limb-darkening effect of the star with a linear approximation. We derived the limb darkening coefficients for each possible temperature of the stellar surface (and spots) and different wavelengths using ExoTETHyS package \citep{2020AJ....159...75M}.\\
To retrieve the spots properties from our data we used the  nested sampling algorithm MultiNest whose priors are shown in Table \ref{PRIORS}, setting 3000 live points to achieve high precision in the retrieval. It should be noted that spot temperature's prior is set with a maximum temperature exceeding that estimated for the unspotted surface to include the possibility of retrieving a facula instead of a spot.\\
In order to fit the data, our forward model simulates the light curve of the star for each spot configuration tested, calculating at each temporal step the portion of the spotted and unspotted stellar surface visible from Earth. Using these informations it estimates the contributions to the total stellar flux of the spotted and of the unspotted visible surface of the star using the method developed in \cite{2021MNRAS.501.1733C}, then it rotates the star of a given time step and uses this loop to calculate the lightcurve of the star. To estimate the fluxes of spots and stellar surface in the chosen range of temperatures (see Table \ref{PRIORS}) we used Phoenix models \citep{2012A&A...546A..14C,2013A&A...553A...6H} and we applied the limb darkening coefficients derived by ExoTETHyS \citep{2020AJ....159...75M}. Finally, MultiNest calculates the fit and Bayesian evidence to estimate the goodness of the fit and in the end the parameters of our model.

\section{Results and discussion}

\subsection{One band vs Multiband approach}

Figure \ref{Old} shows some examples of results obtained from the analysis of single photometric band. It is clear that a degeneracy exists among the latitude of the spots, their radius, and their temperature, as we observe only the emission contrast between the spots and the stellar surface. These degeneracies lead to high uncertainties in spots parameters: in particular, we found a temperature error bar of several hundreds of K, while the latitude spans between 13° and 64° in the R band.
By assuming that the stellar spots and the photosphere emit as star surfaces at different temperatures, the emission contrast should change at different wavelengths, producing a different flux variation across different spectral bands. 
Consequently, through the analysis of simultaneous multiband photometric observations, we could be able to estimate the spots' temperature, thereby breaking or at least reducing the degeneracy between the radius, temperature and latitude of the spots. Therefore, we performed a new retrieval procedure imposing for each 3 days sequence that all the bands share the same spot parameters, but different flux scales.
We applied our model assuming the simultaneous presence of one, two, three, or four spots on the entire stellar surface. However, due to our temporal coverage and sampling, each model with more than one spot resulted in overfitting of our data, so we chose to model only one spot per dataset.
Examples of the resulting corner plots are shown in Figure \ref{CORN1}, \ref{CORN2} and \ref{CORN3} and the retrieved spot temperatures are shown in Table \ref{Temperatures}. In comparison to the single band analysis, we found a more limited interval for spots temperature, with a decrease of the spot temperature uncertainty by a factor of two and a strong reduction of latitude uncertainties. 
The constraints on the spots parameters are typically wider when B filter data are not available, like in the observations of December 11-12-13rd, 2021 and February 21-22-23rd, 2022.
Finally, we checked for the consistency of the spot positions and geometry along each observing run, plotting all the configurations of the spots with weights (of MultiNest algorithm) over the 0.85 quantiles. The above selection guarantees that only high likelihood values are selected and, consequently, shown.

For each observing run, we found a good compatibility of the radius and position of the retrieved spot, even if with some hints of evolution on a timescale similar to the rotational period of the star as we can see comparing the retrieved spots in different 3 days sequences: in each one we had imposed the absence of evolution for the spot, but we can see how the same spot changes between different sequences of days in the same observing run. An example is shown in Figure \ref{Geometry} for the best observing run (February 2021) results. The spot configurations here shown correspond to the same time of the star rotation as obtained using data by each 3 days-sequence of this observing run.
Comparing our results with those of \cite{2023arXiv231214269D}, obtained through a CCF analysis of V1298 Tau's spectra with the hypothesis of non-emitting black spots, our method seems to be sensible only to bigger spots. Our retrieved spots' projected filling factors range at their peak between 0.015 (February 2021) and 0.20 (December 2021 and February 2022), with respect to the distribution of filling factors between 0 and 0.15 found showed in Figure 13 from \cite{2023arXiv231214269D}. Our retrieved projected filling factors are higher than those retrieved through spot CCF, but still compatible. Moreover, our big spots are in the same range of latitude found for big spots using CCF. Our higher retrieved projected filling factors could be explained by our not-optimal time coverage that doesn't allow us to distinguish more than one active areas on the star. The retrieved filling factor is also compatible with results from \citep{2020ApJ...893...67M}, even if it is larger for the last set of data (February 2022).\\

The retrieved spot temperature intervals for each observing run are shown in Table \ref{TEMPSET}.
It is quite evident a decrease of the retrieved spots temperature between the first period of observations and the other two. A possible scenario is that in our approach we are measuring an effective temperature of the inhomogeneities of the stellar surface, which are more complex than simple spots and what we retrieve is in fact the whole active region of the star with properties that result from the average effect of little surface features (like spots and faculae). Then our effective temperature should be sensitive, for example, to a change in the ratio between the filling factor of faculae and spots on the star.  

\subsection{Interpretation of "spot temperature"}
In order to verify the hypothesis that our analysis retrieves the properties of the complex dominant active region, we analyzed the spectroscopic activity indicators of V1298 Tau to get some clues about a possible change of the stellar activity. In particular, we analyzed HARPS-N \citep{2012SPIE.8446E..1VC} data acquired in a span of 4 years (2018-2023) using the method described in \cite{2020A&A...642A..53D}, looking for temporal variability of activity indicators such as CaII H and CaII K (Figure \ref{CaHK}) and $\mathrm{H \alpha}$ (Figure \ref{Halpha}). Vertical lines in the plots mark the periods of our observations. 
In Figure \ref{Halpha} we can see a linear trend for the measured intensity that leads this line from absorption to emission, while in Figure \ref{CaHK} we can see an increase in the scatter of CaII H and K line intensities along the time.
These effects point towards a change in activity between the first and the second period of observations, with an increase of the ratio between $\mathrm{H \alpha}$ and CaII H + CaII K line intensities (Figure \ref{HaCa}). Our photometric observations support that the reason for this evolution could be a change in the ratio of stellar surface covered by spots and faculae.

\subsection{Validation of the model: solar case}

For further validation, we tested the model using Sun observations, where photometric data can be directly compared with spatially resolved active regions present on its surface. We used the multiband photometric data from VIRGO-SPM \citep{1999MNRAS.309..761C}, the three channels sun photometer (SPM) of the experiment VIRGO \citep[Variability of solar IRradiance and Gravity Oscillations]{1997SoPh..170....1F}, an experiment of the space mission SOHO \citep[Solar and Heliospheric Observatory]{1995SoPh..162....1D}.
VIRGO-SPM observed the Sun from April 11th 1996 until March 30th 2014 with an image every 60 seconds in three narrow bands at 402, 500, and 862 nm with a bandwidth of about 5 nm. \\
In particular, we focused on the rotational period starting from 12/30/2013, during which the Sun exhibited a higher level of activity. This was evident through the presence of two large active regions separated by approximately 180° in longitude.
We analyzed the most active one fitting half period. Subsequently, we rebinned the data to obtain two data points per day (which is reasonable for observations of a long-period star like our Sun), and then applied our method to all three bands simultaneously, assuming one dominant spot.
As prior for the retrieval model we imposed a spot radius between 0 and 0.25 and a spot temperature between 3800 and 5900 K to include also the possibility of retrieval of faculae instead of spots. The fit results for the first half-period analyzed are shown in Figure \ref{CORNERSUN}. \\
Figure \ref{SUN} shows the comparison between the simulation of the retrieved solar active region (left panel) and the actual contemporary image of the Sun (right panel), with its real spots, taken from Helioseismic and Magnetic Imager \citep[HMI]{2012SoPh..275..229S}, while the bottom panel of the figure shows how our model fit the observed solar data in the three photometric bands.
The comparison between the two upper panels of Figure \ref{SUN} shows that we were able to constrain with great precision the position of the active regions of the Sun breaking the degeneracy between their latitude and radius, while Figure \ref{CORNERSUN} shows that the degeneracy between active region's latitude and temperature is also small and confined to a narrow parameters range. Moreover, we found that the retrieved radius encloses not only the main big spot but also all the smaller spots, penumbra areas, faculae and some of the photosphere inside and around them, leading to very little contrast between the retrieved active region's temperature and the photospheric temperature, confirming our initial hypothesis that our procedure retrieves the average temperature of the entire active region of the star.

\section{Summary and Conclusions}

In conclusion, modelling multiband photometry and assuming a single spot-dominated surface, we measured an effective temperature of the active regions of V1298 Tau which is probably a weighted average of the temperature of the spots, faculae and photosphere around them. Multiband photometry allows us to estimate this property, with the bluest photometric bands that provide the strongest constraint.
We may reduce the degeneracy between radius, latitude and temperature of the active regions of the star, thanks to simultaneous multiband data, especially retrieving correctly its geometric properties, as verified on the solar case. 
More bands, especially bluer bands, and a denser temporal coverage, likely, would help in reducing the temperature uncertainties and in distinguishing spots and other features of the surface and better constraining their positions. 
Finally, we developed a method to characterize surface inhomogeneities of quite active and young stars that has the advantage of using little and medium-size telescopes for few days. 
Our results strongly suggest that the evolution of the effective temperature of the active regions may reflect the evolution of stellar activity, indicating the needs for a monitoring of the star on various timescales.
In conclusion, we suggest that this method, together with ground-based high-resolution spectroscopy and photometric observations from space, could improve our comprehension of active stars. \\

\begin{acknowledgements}
      The authors thank the anonymous reviewer for his/her very useful comments and suggestions.
      The authors acknowledge the support of Osservatorio Polifunzionale del Chianti for the acquisition of the data.\\
      The authors acknowledge the support of the ASI-INAF agreement 2021-5-HH.0. Part of the research activities described in this paper were carried out with contribution of the Next Generation EU funds within the National Recovery and Resilience Plan (PNRR), Mission 4 - Education and Research, Component 2 - From Research to Business (M4C2), Investment Line 3.1 - Strengthening and creation of Research Infrastructures, Project IR0000034 – “STILES - Strengthening the Italian Leadership in ELT and SKA”.
      AP and GM acknowledge the support of grant n. 2022J7ZFRA - Exo-planetary Cloudy Atmospheres and Stellar High energy (Exo-CASH) funded by MUR - PRIN 2022.
      AB and AP acknowledge the support from the INAF Minigrant of the RSN-2 nr. 16 "SpAcES: Spotting the Activity of Exoplanet hosting Stars" according to the INAF Fundamental Astrophysics funding scheme.
\end{acknowledgements}


\bibliographystyle{aa}
\bibliography{V1298_SPOT} 

\begin{thebibliography}{49}
\expandafter\ifx\csname natexlab\endcsname\relax\def\natexlab#1{#1}\fi

\bibitem[{{Agol} {et~al.}(2010){Agol}, {Cowan}, {Knutson}, {Deming}, {Steffen}, {Henry}, \& {Charbonneau}}]{2010ApJ...721.1861A}
{Agol}, E., {Cowan}, N.~B., {Knutson}, H.~A., {et~al.} 2010, \apj, 721, 1861

\bibitem[{{Baliunas} {et~al.}(1995){Baliunas}, {Donahue}, {Soon}, {Horne}, {Frazer}, {Woodard-Eklund}, {Bradford}, {Rao}, {Wilson}, {Zhang}, {Bennett}, {Briggs}, {Carroll}, {Duncan}, {Figueroa}, {Lanning}, {Misch}, {Mueller}, {Noyes}, {Poppe}, {Porter}, {Robinson}, {Russell}, {Shelton}, {Soyumer}, {Vaughan}, \& {Whitney}}]{1995ApJ...438..269B}
{Baliunas}, S.~L., {Donahue}, R.~A., {Soon}, W.~H., {et~al.} 1995, \apj, 438, 269

\bibitem[{{Ballerini} {et~al.}(2012){Ballerini}, {Micela}, {Lanza}, \& {Pagano}}]{2012A&A...539A.140B}
{Ballerini}, P., {Micela}, G., {Lanza}, A.~F., \& {Pagano}, I. 2012, \aap, 539, A140

\bibitem[{{Berdyugina}(2005)}]{2005LRSP....2....8B}
{Berdyugina}, S.~V. 2005, Living Reviews in Solar Physics, 2, 8

\bibitem[{{Berta} {et~al.}(2011){Berta}, {Charbonneau}, {Bean}, {Irwin}, {Burke}, {D{\'e}sert}, {Nutzman}, \& {Falco}}]{2011ApJ...736...12B}
{Berta}, Z.~K., {Charbonneau}, D., {Bean}, J., {et~al.} 2011, \apj, 736, 12

\bibitem[{{Buchner} {et~al.}(2014){Buchner}, {Georgakakis}, {Nandra}, {Hsu}, {Rangel}, {Brightman}, {Merloni}, {Salvato}, {Donley}, \& {Kocevski}}]{2014A&A...564A.125B}
{Buchner}, J., {Georgakakis}, A., {Nandra}, K., {et~al.} 2014, \aap, 564, A125

\bibitem[{{Carleo} {et~al.}(2020){Carleo}, {Malavolta}, {Lanza}, {Damasso}, {Desidera}, {Borsa}, {Mallonn}, {Pinamonti}, {Gratton}, {Alei}, {Benatti}, {Mancini}, {Maldonado}, {Biazzo}, {Esposito}, {Frustagli}, {Gonz{\'a}lez-{\'A}lvarez}, {Micela}, {Scandariato}, {Sozzetti}, {Affer}, {Bignamini}, {Bonomo}, {Claudi}, {Cosentino}, {Covino}, {Fiorenzano}, {Giacobbe}, {Harutyunyan}, {Leto}, {Maggio}, {Molinari}, {Nascimbeni}, {Pagano}, {Pedani}, {Piotto}, {Poretti}, {Rainer}, {Redfield}, {Baffa}, {Baruffolo}, {Buchschacher}, {Billotti}, {Cecconi}, {Falcini}, {Fantinel}, {Fini}, {Galli}, {Ghedina}, {Ghinassi}, {Giani}, {Gonzalez}, {Gonzalez}, {Guerra}, {Hernandez Diaz}, {Hernandez}, {Iuzzolino}, {Lodi}, {Oliva}, {Origlia}, {Perez Ventura}, {Puglisi}, {Riverol}, {Riverol}, {San Juan}, {Sanna}, {Scuderi}, {Seemann}, {Sozzi}, \& {Tozzi}}]{2020A&A...638A...5C}
{Carleo}, I., {Malavolta}, L., {Lanza}, A.~F., {et~al.} 2020, \aap, 638, A5

\bibitem[{{Changeat} {et~al.}(2020){Changeat}, {Edwards}, {Al-Refaie}, {Morvan}, {Tsiaras}, {Waldmann}, \& {Tinetti}}]{2020AJ....160..260C}
{Changeat}, Q., {Edwards}, B., {Al-Refaie}, A.~F., {et~al.} 2020, \aj, 160, 260

\bibitem[{{Chaplin} \& {Appourchaux}(1999)}]{1999MNRAS.309..761C}
{Chaplin}, W.~J. \& {Appourchaux}, T. 1999, \mnras, 309, 761

\bibitem[{{Claret} {et~al.}(2012){Claret}, {Hauschildt}, \& {Witte}}]{2012A&A...546A..14C}
{Claret}, A., {Hauschildt}, P.~H., \& {Witte}, S. 2012, \aap, 546, A14

\bibitem[{{Collins} {et~al.}(2017){Collins}, {Kielkopf}, {Stassun}, \& {Hessman}}]{2017AJ....153...77C}
{Collins}, K.~A., {Kielkopf}, J.~F., {Stassun}, K.~G., \& {Hessman}, F.~V. 2017, \aj, 153, 77

\bibitem[{{Cosentino} {et~al.}(2012){Cosentino}, {Lovis}, {Pepe}, {Collier Cameron}, {Latham}, {Molinari}, {Udry}, {Bezawada}, {Black}, {Born}, {Buchschacher}, {Charbonneau}, {Figueira}, {Fleury}, {Galli}, {Gallie}, {Gao}, {Ghedina}, {Gonzalez}, {Gonzalez}, {Guerra}, {Henry}, {Horne}, {Hughes}, {Kelly}, {Lodi}, {Lunney}, {Maire}, {Mayor}, {Micela}, {Ordway}, {Peacock}, {Phillips}, {Piotto}, {Pollacco}, {Queloz}, {Rice}, {Riverol}, {Riverol}, {San Juan}, {Sasselov}, {Segransan}, {Sozzetti}, {Sosnowska}, {Stobie}, {Szentgyorgyi}, {Vick}, \& {Weber}}]{2012SPIE.8446E..1VC}
{Cosentino}, R., {Lovis}, C., {Pepe}, F., {et~al.} 2012, in Society of Photo-Optical Instrumentation Engineers (SPIE) Conference Series, Vol. 8446, Ground-based and Airborne Instrumentation for Astronomy IV, ed. I.~S. {McLean}, S.~K. {Ramsay}, \& H.~{Takami}, 84461V

\bibitem[{{Cracchiolo} {et~al.}(2021{\natexlab{a}}){Cracchiolo}, {Micela}, {Morello}, \& {Peres}}]{2021MNRAS.507.6118C}
{Cracchiolo}, G., {Micela}, G., {Morello}, G., \& {Peres}, G. 2021{\natexlab{a}}, \mnras, 507, 6118

\bibitem[{{Cracchiolo} {et~al.}(2021{\natexlab{b}}){Cracchiolo}, {Micela}, \& {Peres}}]{2021MNRAS.501.1733C}
{Cracchiolo}, G., {Micela}, G., \& {Peres}, G. 2021{\natexlab{b}}, \mnras, 501, 1733

\bibitem[{{Czesla} {et~al.}(2009){Czesla}, {Huber}, {Wolter}, {Schr{\"o}ter}, \& {Schmitt}}]{2009A&A...505.1277C}
{Czesla}, S., {Huber}, K.~F., {Wolter}, U., {Schr{\"o}ter}, S., \& {Schmitt}, J.~H.~M.~M. 2009, \aap, 505, 1277

\bibitem[{{Damasso} {et~al.}(2023){Damasso}, {Locci}, {Benatti}, {Maggio}, {Nardiello}, {Baratella}, {Biazzo}, {Bonomo}, {Desidera}, {D'Orazi}, {Mallonn}, {Lanza}, {Sozzetti}, {Marzari}, {Borsa}, {Maldonado}, {Mancini}, {Poretti}, {Scandariato}, {Bignamini}, {Borsato}, {Capuzzo Dolcetta}, {Cecconi}, {Claudi}, {Cosentino}, {Covino}, {Fiorenzano}, {Harutyunyan}, {Mann}, {Micela}, {Molinari}, {Molinaro}, {Pagano}, {Pedani}, {Pinamonti}, {Piotto}, \& {Stoev}}]{2023A&A...672A.126D}
{Damasso}, M., {Locci}, D., {Benatti}, S., {et~al.} 2023, \aap, 672, A126

\bibitem[{{David} {et~al.}(2019{\natexlab{a}}){David}, {Cody}, {Hedges}, {Mamajek}, {Hillenbrand}, {Ciardi}, {Beichman}, {Petigura}, {Fulton}, {Isaacson}, {Howard}, {Gagn{\'e}}, {Saunders}, {Rebull}, {Stauffer}, {Vasisht}, \& {Hinkley}}]{2019AJ....158...79D}
{David}, T.~J., {Cody}, A.~M., {Hedges}, C.~L., {et~al.} 2019{\natexlab{a}}, \aj, 158, 79

\bibitem[{{David} {et~al.}(2016){David}, {Conroy}, {Hillenbrand}, {Stassun}, {Stauffer}, {Rebull}, {Cody}, {Isaacson}, {Howard}, \& {Aigrain}}]{2016AJ....151..112D}
{David}, T.~J., {Conroy}, K.~E., {Hillenbrand}, L.~A., {et~al.} 2016, \aj, 151, 112

\bibitem[{{David} {et~al.}(2019{\natexlab{b}}){David}, {Petigura}, {Luger}, {Foreman-Mackey}, {Livingston}, {Mamajek}, \& {Hillenbrand}}]{2019ApJ...885L..12D}
{David}, T.~J., {Petigura}, E.~A., {Luger}, R., {et~al.} 2019{\natexlab{b}}, \apjl, 885, L12

\bibitem[{{D{\'e}sert} {et~al.}(2011){D{\'e}sert}, {Sing}, {Vidal-Madjar}, {H{\'e}brard}, {Ehrenreich}, {Lecavelier Des Etangs}, {Parmentier}, {Ferlet}, \& {Henry}}]{2011A&A...526A..12D}
{D{\'e}sert}, J.~M., {Sing}, D., {Vidal-Madjar}, A., {et~al.} 2011, \aap, 526, A12

\bibitem[{{Di Maio} {et~al.}(2020){Di Maio}, {Argiroffi}, {Micela}, {Benatti}, {Lanza}, {Scandariato}, {Maldonado}, {Maggio}, {Affer}, \& {Claudi}}]{2020A&A...642A..53D}
{Di Maio}, C., {Argiroffi}, C., {Micela}, G., {et~al.} 2020, \aap, 642, A53

\bibitem[{{Di Maio} {et~al.}(2023){Di Maio}, {Petralia}, {Micela}, {Lanza}, {Rainer}, {Malavolta}, {Benatti}, {Affer}, {Maldonado}, {Colombo}, {Damasso}, {Maggio}, {Biazzo}, {Bignamini}, {Borsa}, {Boschin}, {Cabona}, {Cecconi}, {Claudi}, {Covino}, {Di Fabrizio}, {Gratton}, {Lorenzi}, {Mancini}, {Messina}, {Molinari}, {Molinaro}, {Nardiello}, {Poretti}, \& {Sozzetti}}]{2023arXiv231214269D}
{Di Maio}, C., {Petralia}, A., {Micela}, G., {et~al.} 2023, arXiv e-prints, arXiv:2312.14269

\bibitem[{{Domingo} {et~al.}(1995){Domingo}, {Fleck}, \& {Poland}}]{1995SoPh..162....1D}
{Domingo}, V., {Fleck}, B., \& {Poland}, A.~I. 1995, \solphys, 162, 1

\bibitem[{{Feinstein} {et~al.}(2022){Feinstein}, {David}, {Montet}, {Foreman-Mackey}, {Livingston}, \& {Mann}}]{2022ApJ...925L...2F}
{Feinstein}, A.~D., {David}, T.~J., {Montet}, B.~T., {et~al.} 2022, \apjl, 925, L2

\bibitem[{{Feroz} {et~al.}(2009){Feroz}, {Hobson}, \& {Bridges}}]{2009MNRAS.398.1601F}
{Feroz}, F., {Hobson}, M.~P., \& {Bridges}, M. 2009, \mnras, 398, 1601

\bibitem[{{Fr{\"o}hlich} {et~al.}(1997){Fr{\"o}hlich}, {Andersen}, {Appourchaux}, {Berthomieu}, {Crommelynck}, {Domingo}, {Fichot}, {Finsterle}, {G{\'o}mez}, {Gough}, {Jim{\'e}nez}, {Leifsen}, {Lombaerts}, {Pap}, {Provost}, {Roca Cort{\'e}s}, {Romero}, {Roth}, {Sekii}, {Telljohann}, {Toutain}, \& {Wehrli}}]{1997SoPh..170....1F}
{Fr{\"o}hlich}, C., {Andersen}, B.~N., {Appourchaux}, T., {et~al.} 1997, \solphys, 170, 1

\bibitem[{{Greene} {et~al.}(2016){Greene}, {Line}, {Montero}, {Fortney}, {Lustig-Yaeger}, \& {Luther}}]{2016ApJ...817...17G}
{Greene}, T.~P., {Line}, M.~R., {Montero}, C., {et~al.} 2016, \apj, 817, 17

\bibitem[{{Husser} {et~al.}(2013){Husser}, {Wende-von Berg}, {Dreizler}, {Homeier}, {Reiners}, {Barman}, \& {Hauschildt}}]{2013A&A...553A...6H}
{Husser}, T.~O., {Wende-von Berg}, S., {Dreizler}, S., {et~al.} 2013, \aap, 553, A6

\bibitem[{{Latham} {et~al.}(2011){Latham}, {Rowe}, {Quinn}, {Batalha}, {Borucki}, {Brown}, {Bryson}, {Buchhave}, {Caldwell}, {Carter}, {Christiansen}, {Ciardi}, {Cochran}, {Dunham}, {Fabrycky}, {Ford}, {Gautier}, {Gilliland}, {Holman}, {Howell}, {Ibrahim}, {Isaacson}, {Jenkins}, {Koch}, {Lissauer}, {Marcy}, {Quintana}, {Ragozzine}, {Sasselov}, {Shporer}, {Steffen}, {Welsh}, \& {Wohler}}]{2011ApJ...732L..24L}
{Latham}, D.~W., {Rowe}, J.~F., {Quinn}, S.~N., {et~al.} 2011, \apjl, 732, L24

\bibitem[{{Luger} {et~al.}(2021){Luger}, {Foreman-Mackey}, {Hedges}, \& {Hogg}}]{2021AJ....162..123L}
{Luger}, R., {Foreman-Mackey}, D., {Hedges}, C., \& {Hogg}, D.~W. 2021, \aj, 162, 123

\bibitem[{{Mantovan} {et~al.}(2024){Mantovan}, {Malavolta}, {Desidera}, {Zingales}, {Borsato}, {Piotto}, {Maggio}, {Locci}, {Polychroni}, {Turrini}, {Baratella}, {Biazzo}, {Nardiello}, {Stassun}, {Nascimbeni}, {Benatti}, {John}, {Watkins}, {Bieryla}, {Lissauer}, {Twicken}, {Lanza}, {Winn}, {Messina}, {Montalto}, {Sozzetti}, {Boffin}, {Cheryasov}, {Strakhov}, {Murgas}, {D'Arpa}, {Barkaoui}, {Benni}, {Bignamini}, {Bonomo}, {Borsa}, {Cabona}, {Cameron}, {Claudi}, {Cochran}, {Collins}, {Damasso}, {Dong}, {Endl}, {Fukui}, {F{\H{u}}r{\'e}sz}, {Gandolfi}, {Ghedina}, {Jenkins}, {Kab{\'a}th}, {Latham}, {Lorenzi}, {Luque}, {Maldonado}, {McLeod}, {Molinaro}, {Narita}, {Nowak}, {Orell-Miquel}, {Pall{\'e}}, {Parviainen}, {Pedani}, {Quinn}, {Relles}, {Rowden}, {Scandariato}, {Schwarz}, {Seager}, {Shporer}, {Vanderburg}, \& {Wilson}}]{2024A&A...682A.129M}
{Mantovan}, G., {Malavolta}, L., {Desidera}, S., {et~al.} 2024, \aap, 682, A129

\bibitem[{{Micela}(2015)}]{2015ExA....40..723M}
{Micela}, G. 2015, Experimental Astronomy, 40, 723

\bibitem[{{Morello} {et~al.}(2020){Morello}, {Claret}, {Martin-Lagarde}, {Cossou}, {Tsiaras}, \& {Lagage}}]{2020AJ....159...75M}
{Morello}, G., {Claret}, A., {Martin-Lagarde}, M., {et~al.} 2020, \aj, 159, 75

\bibitem[{{Morris}(2020)}]{2020ApJ...893...67M}
{Morris}, B.~M. 2020, \apj, 893, 67

\bibitem[{{Pallavicini} {et~al.}(1981){Pallavicini}, {Golub}, {Rosner}, {Vaiana}, {Ayres}, \& {Linsky}}]{1981ApJ...248..279P}
{Pallavicini}, R., {Golub}, L., {Rosner}, R., {et~al.} 1981, \apj, 248, 279

\bibitem[{{Pizzolato} {et~al.}(2003){Pizzolato}, {Maggio}, {Micela}, {Sciortino}, \& {Ventura}}]{2003A&A...397..147P}
{Pizzolato}, N., {Maggio}, A., {Micela}, G., {Sciortino}, S., \& {Ventura}, P. 2003, \aap, 397, 147

\bibitem[{{Plavchan} {et~al.}(2020){Plavchan}, {Barclay}, {Gagn{\'e}}, {Gao}, {Cale}, {Matzko}, {Dragomir}, {Quinn}, {Feliz}, {Stassun}, {Crossfield}, {Berardo}, {Latham}, {Tieu}, {Anglada-Escud{\'e}}, {Ricker}, {Vanderspek}, {Seager}, {Winn}, {Jenkins}, {Rinehart}, {Krishnamurthy}, {Dynes}, {Doty}, {Adams}, {Afanasev}, {Beichman}, {Bottom}, {Bowler}, {Brinkworth}, {Brown}, {Cancino}, {Ciardi}, {Clampin}, {Clark}, {Collins}, {Davison}, {Foreman-Mackey}, {Furlan}, {Gaidos}, {Geneser}, {Giddens}, {Gilbert}, {Hall}, {Hellier}, {Henry}, {Horner}, {Howard}, {Huang}, {Huber}, {Kane}, {Kenworthy}, {Kielkopf}, {Kipping}, {Klenke}, {Kruse}, {Latouf}, {Lowrance}, {Mennesson}, {Mengel}, {Mills}, {Morton}, {Narita}, {Newton}, {Nishimoto}, {Okumura}, {Palle}, {Pepper}, {Quintana}, {Roberge}, {Roccatagliata}, {Schlieder}, {Tanner}, {Teske}, {Tinney}, {Vanderburg}, {von Braun}, {Walp}, {Wang}, {Wang}, {Weigand}, {White}, {Wittenmyer}, {Wright}, {Youngblood}, {Zhang}, \& {Zilberman}}]{2020Natur.582..497P}
{Plavchan}, P., {Barclay}, T., {Gagn{\'e}}, J., {et~al.} 2020, \nat, 582, 497

\bibitem[{{Pont} {et~al.}(2008){Pont}, {Knutson}, {Gilliland}, {Moutou}, \& {Charbonneau}}]{2008MNRAS.385..109P}
{Pont}, F., {Knutson}, H., {Gilliland}, R.~L., {Moutou}, C., \& {Charbonneau}, D. 2008, \mnras, 385, 109

\bibitem[{{Rizzuto} {et~al.}(2020){Rizzuto}, {Newton}, {Mann}, {Tofflemire}, {Vanderburg}, {Kraus}, {Wood}, {Quinn}, {Zhou}, {Thao}, {Law}, {Ziegler}, \& {Brice{\~n}o}}]{2020AJ....160...33R}
{Rizzuto}, A.~C., {Newton}, E.~R., {Mann}, A.~W., {et~al.} 2020, \aj, 160, 33

\bibitem[{{Scandariato} \& {Micela}(2015)}]{2015ExA....40..711S}
{Scandariato}, G. \& {Micela}, G. 2015, Experimental Astronomy, 40, 711

\bibitem[{{Schou} {et~al.}(2012){Schou}, {Scherrer}, {Bush}, {Wachter}, {Couvidat}, {Rabello-Soares}, {Bogart}, {Hoeksema}, {Liu}, {Duvall}, {Akin}, {Allard}, {Miles}, {Rairden}, {Shine}, {Tarbell}, {Title}, {Wolfson}, {Elmore}, {Norton}, \& {Tomczyk}}]{2012SoPh..275..229S}
{Schou}, J., {Scherrer}, P.~H., {Bush}, R.~I., {et~al.} 2012, \solphys, 275, 229

\bibitem[{{Sing} {et~al.}(2011{\natexlab{a}}){Sing}, {D{\'e}sert}, {Fortney}, {Lecavelier Des Etangs}, {Ballester}, {Cepa}, {Ehrenreich}, {L{\'o}pez-Morales}, {Pont}, {Shabram}, \& {Vidal-Madjar}}]{2011A&A...527A..73S}
{Sing}, D.~K., {D{\'e}sert}, J.~M., {Fortney}, J.~J., {et~al.} 2011{\natexlab{a}}, \aap, 527, A73

\bibitem[{{Sing} {et~al.}(2009){Sing}, {D{\'e}sert}, {Lecavelier Des Etangs}, {Ballester}, {Vidal-Madjar}, {Parmentier}, {Hebrard}, \& {Henry}}]{2009A&A...505..891S}
{Sing}, D.~K., {D{\'e}sert}, J.~M., {Lecavelier Des Etangs}, A., {et~al.} 2009, \aap, 505, 891

\bibitem[{{Sing} {et~al.}(2011{\natexlab{b}}){Sing}, {Pont}, {Aigrain}, {Charbonneau}, {D{\'e}sert}, {Gibson}, {Gilliland}, {Hayek}, {Henry}, {Knutson}, {Lecavelier Des Etangs}, {Mazeh}, \& {Shporer}}]{2011MNRAS.416.1443S}
{Sing}, D.~K., {Pont}, F., {Aigrain}, S., {et~al.} 2011{\natexlab{b}}, \mnras, 416, 1443

\bibitem[{{Solanki}(2003)}]{2003A&ARv..11..153S}
{Solanki}, S.~K. 2003, \aapr, 11, 153

\bibitem[{{Su{\'a}rez Mascare{\~n}o} {et~al.}(2021){Su{\'a}rez Mascare{\~n}o}, {Damasso}, {Lodieu}, {Sozzetti}, {B{\'e}jar}, {Benatti}, {Zapatero Osorio}, {Micela}, {Rebolo}, {Desidera}, {Murgas}, {Claudi}, {Gonz{\'a}lez Hern{\'a}ndez}, {Malavolta}, {del Burgo}, {D'Orazi}, {Amado}, {Locci}, {Tabernero}, {Marzari}, {Aguado}, {Turrini}, {Cardona Guill{\'e}n}, {Toledo-Padr{\'o}n}, {Maggio}, {Aceituno}, {Bauer}, {Caballero}, {Chinchilla}, {Esparza-Borges}, {Gonz{\'a}lez-{\'A}lvarez}, {Granzer}, {Luque}, {Mart{\'\i}n}, {Nowak}, {Oshagh}, {Pall{\'e}}, {Parviainen}, {Quirrenbach}, {Reiners}, {Ribas}, {Strassmeier}, {Weber}, \& {Mallonn}}]{2022NatAs...6..232S}
{Su{\'a}rez Mascare{\~n}o}, A., {Damasso}, M., {Lodieu}, N., {et~al.} 2021, Nature Astronomy, 6, 232

\bibitem[{{Tinetti} {et~al.}(2018){Tinetti}, {Drossart}, {Eccleston}, {Hartogh}, {Heske}, {Leconte}, {Micela}, {Ollivier}, {Pilbratt}, {Puig}, {Turrini}, {Vandenbussche}, {Wolkenberg}, {Beaulieu}, {Buchave}, {Ferus}, {Griffin}, {Guedel}, {Justtanont}, {Lagage}, {Machado}, {Malaguti}, {Min}, {N{\o}rgaard-Nielsen}, {Rataj}, {Ray}, {Ribas}, {Swain}, {Szabo}, {Werner}, {Barstow}, {Burleigh}, {Cho}, {Coud{\'e} du Foresto}, {Coustenis}, {Decin}, {Encrenaz}, {Galand}, {Gillon}, {Helled}, {Morales}, {Garc{\'\i}a Mu{\~n}oz}, {Moneti}, {Pagano}, {Pascale}, {Piccioni}, {Pinfield}, {Sarkar}, {Selsis}, {Tennyson}, {Triaud}, {Venot}, {Waldmann}, {Waltham}, {Wright}, {Amiaux}, {Augu{\`e}res}, {Berth{\'e}}, {Bezawada}, {Bishop}, {Bowles}, {Coffey}, {Colom{\'e}}, {Crook}, {Crouzet}, {Da Peppo}, {Sanz}, {Focardi}, {Frericks}, {Hunt}, {Kohley}, {Middleton}, {Morgante}, {Ottensamer}, {Pace}, {Pearson}, {Stamper}, {Symonds}, {Rengel}, {Renotte}, {Ade}, {Affer}, {Alard}, {Allard}, {Altieri}, {Andr{\'e}}, {Arena}, {Argyriou},
  {Aylward}, {Baccani}, {Bakos}, {Banaszkiewicz}, {Barlow}, {Batista}, {Bellucci}, {Benatti}, {Bernardi}, {B{\'e}zard}, {Blecka}, {Bolmont}, {Bonfond}, {Bonito}, {Bonomo}, {Brucato}, {Brun}, {Bryson}, {Bujwan}, {Casewell}, {Charnay}, {Pestellini}, {Chen}, {Ciaravella}, {Claudi}, {Cl{\'e}dassou}, {Damasso}, {Damiano}, {Danielski}, {Deroo}, {Di Giorgio}, {Dominik}, {Doublier}, {Doyle}, {Doyon}, {Drummond}, {Duong}, {Eales}, {Edwards}, {Farina}, {Flaccomio}, {Fletcher}, {Forget}, {Fossey}, {Fr{\"a}nz}, {Fujii}, {Garc{\'\i}a-Piquer}, {Gear}, {Geoffray}, {G{\'e}rard}, {Gesa}, {Gomez}, {Graczyk}, {Griffith}, {Grodent}, {Guarcello}, {Gustin}, {Hamano}, {Hargrave}, {Hello}, {Heng}, {Herrero}, {Hornstrup}, {Hubert}, {Ida}, {Ikoma}, {Iro}, {Irwin}, {Jarchow}, {Jaubert}, {Jones}, {Julien}, {Kameda}, {Kerschbaum}, {Kervella}, {Koskinen}, {Krijger}, {Krupp}, {Lafarga}, {Landini}, {Lellouch}, {Leto}, {Luntzer}, {Rank-L{\"u}ftinger}, {Maggio}, {Maldonado}, {Maillard}, {Mall}, {Marquette}, {Mathis}, {Maxted}, {Matsuo},
  {Medvedev}, {Miguel}, {Minier}, {Morello}, {Mura}, {Narita}, {Nascimbeni}, {Nguyen Tong}, {Noce}, {Oliva}, {Palle}, {Palmer}, {Pancrazzi}, {Papageorgiou}, {Parmentier}, {Perger}, {Petralia}, {Pezzuto}, {Pierrehumbert}, {Pillitteri}, {Piotto}, {Pisano}, {Prisinzano}, {Radioti}, {R{\'e}ess}, {Rezac}, {Rocchetto}, {Rosich}, {Sanna}, {Santerne}, {Savini}, {Scandariato}, {Sicardy}, {Sierra}, {Sindoni}, {Skup}, {Snellen}, {Sobiecki}, {Soret}, {Sozzetti}, {Stiepen}, {Strugarek}, {Taylor}, {Taylor}, {Terenzi}, {Tessenyi}, {Tsiaras}, {Tucker}, {Valencia}, {Vasisht}, {Vazan}, {Vilardell}, {Vinatier}, {Viti}, {Waters}, {Wawer}, {Wawrzaszek}, {Whitworth}, {Yung}, {Yurchenko}, {Zapatero Osorio}, {Zellem}, {Zingales}, \& {Zwart}}]{2018ExA....46..135T}
{Tinetti}, G., {Drossart}, P., {Eccleston}, P., {et~al.} 2018, Experimental Astronomy, 46, 135

\bibitem[{{Tsiaras} {et~al.}(2018){Tsiaras}, {Waldmann}, {Zingales}, {Rocchetto}, {Morello}, {Damiano}, {Karpouzas}, {Tinetti}, {McKemmish}, {Tennyson}, \& {Yurchenko}}]{2018AJ....155..156T}
{Tsiaras}, A., {Waldmann}, I.~P., {Zingales}, T., {et~al.} 2018, \aj, 155, 156

\bibitem[{{Walter} \& {Bowyer}(1981)}]{1981ApJ...245..671W}
{Walter}, F.~M. \& {Bowyer}, S. 1981, \apj, 245, 671

\end{thebibliography}

\end{document}